\documentclass[preprint]{aastex}

\usepackage{graphicx}
\usepackage{longtable}
\usepackage{epstopdf, epsfig}
 \usepackage{epsfig}

\usepackage{amssymb}
\usepackage{amsmath}

\usepackage{multirow}
\usepackage[flushleft]{threeparttable}

\usepackage{lineno}
\usepackage{multirow}
\usepackage{booktabs}
\usepackage{amsbsy}
\usepackage{soul}
\usepackage{color}
\usepackage{units}
\usepackage{hyperref}

\usepackage{float}
\restylefloat{table}

\journal{ApJ}

\begin{document}

\shorttitle{Searches for Neutrino Sources}
\shortauthors{M.~G.~Aartsen et al.}

\title{Searches for Extended and Point-like Neutrino Sources with Four Years of IceCube Data}

\author{
IceCube Collaboration:
M.~G.~Aartsen\altaffilmark{1},
M.~Ackermann\altaffilmark{2},
J.~Adams\altaffilmark{3},
J.~A.~Aguilar\altaffilmark{4},
M.~Ahlers\altaffilmark{5},
M.~Ahrens\altaffilmark{6},
D.~Altmann\altaffilmark{7},
T.~Anderson\altaffilmark{8},
C.~Arguelles\altaffilmark{5},
T.~C.~Arlen\altaffilmark{8}
% ...
}
\altaffiltext{1}{School of Chemistry \& Physics, University of Adelaide, Adelaide SA, 5005 Australia}
\altaffiltext{2}{DESY, D-15735 Zeuthen, Germany}
\altaffiltext{3}{Dept.~of Physics and Astronomy, University of Canterbury, Private Bag 4800, Christchurch, New Zealand}
\altaffiltext{4}{D\'epartement de physique nucl\'eaire et corpusculaire, Universit\'e de Gen\`eve, CH-1211 Gen\`eve, Switzerland}
\altaffiltext{5}{Dept.~of Physics and Wisconsin IceCube Particle Astrophysics Center, University of Wisconsin, Madison, WI 53706, USA}
\altaffiltext{6}{Oskar Klein Centre and Dept.~of Physics, Stockholm University, SE-10691 Stockholm, Sweden}
\altaffiltext{7}{Erlangen Centre for Astroparticle Physics, Friedrich-Alexander-Universit\"at Erlangen-N\"urnberg, D-91058 Erlangen, Germany}
\altaffiltext{8}{Dept.~of Physics, Pennsylvania State University, University Park, PA 16802, USA}
% ...

\begin{abstract}

We present results on searches for point-like sources of neutrinos using four years of IceCube data, including the first year of data from the completed 86-string detector. The total livetime of the combined dataset is 1,373 days. For an E$^{-2}$ spectrum the  median sensitivity at 90\% C.L. is $\sim 10^{-12}$\,TeV$^{-1}$cm$^{-2}$s$^{-1}$ for energies between 1\,TeV$-$1\,PeV in the northern sky and $\sim 10^{-11}$\,TeV$^{-1}$cm$^{-2}$s$^{-1}$ for energies between 100\,TeV $-$ 100\,PeV in the southern sky. The sensitivity has improved from both the additional year of data and the introduction of improved reconstructions compared to previous publications. In addition, we present the first results from an all-sky search for extended sources of neutrinos. We update results of searches for neutrino emission from stacked catalogs of sources, and test five new catalogs; two of Galactic supernova remnants and three of active galactic nuclei. In all cases, the data are compatible with the background-only hypothesis, and upper limits on the 
flux of muon neutrinos are reported for the sources considered.

\end{abstract}


\section{Introduction}
\label{sec1}

Neutrinos have unique properties that can be used to probe diverse astrophysical processes.  Produced in interactions of protons and nuclei with ambient radiation and matter, their low cross-section allows them to travel astronomical distances without experiencing significant absorption.  Unlike charged cosmic rays which change direction as they pass through galactic and intergalactic magnetic fields, neutrinos preserve their directional information as they travel straight from the source to Earth.  Astrophysical neutrinos are also tracers of hadronic interactions, and the identification of these neutrino sources may help to clarify cosmic ray acceleration processes \citep{AnchorMonta,AnchorReview,2008Becker,2002HalzenHooper,2000LearnedMannheim}. Candidate sources for cosmic ray acceleration (and therefore neutrino emission) include Supernova Remnant (SNR) shocks \citep{2002ApJL,MC,MilagroNewHalzenPaper,2006PRD,2011AstroPart}, Active Galactic Nuclei (AGN) jets \citep{2010PRL,2013PRL,MuraseAGN,2011ApJ,1991Stecker,WaxmanBahcall}, Starburst Galaxies \citep{2011ApJStarbursts,LoebWaxman,MuraseStarbursts,2003ApJStarbursts}, and Gamma-Ray Bursts (GRBs) \citep{Guetta,Meszaros,WaxmanBahcallGRB}.

IceCube recently found evidence for a diffuse flux of high-energy astrophysical neutrinos~\citep{HESEPaper,HESE3Year}, observing a $5.7\sigma$ excess of events between $\sim 50$\,\unit{TeV} and 2\,\unit{PeV} deposited within the detector.  The 37 observed events are consistent with an E$^{-2.3}$ neutrino flux at the level of $1.5 \times10^{-11}$\,\unit{TeV$^{-1}$}\unit{cm$^{-2}$}\unit{s$^{-1}$}\unit{sr$^{-1}$} (normalized at 100 \unit{TeV}), with a neutrino flavor ratio of 1:1:1. While these events have established unequivocally that astrophysical neutrinos exist, their sources have not yet been identified.  One challenge is that only $\sim20\%$ of the events in that sample are associated with a high-energy muon which leaves a visible track in the detector. The remaining events without a track have a poor angular resolution of $\sim 15^{\circ}$.

This paper presents the latest results of searches for point sources of astrophysical neutrinos with a sample of track-like events associated with $\nu_{\mu}$ (and some $\nu_{\tau}$) charged current interactions observed by the IceCube Neutrino Observatory. These events have an excellent angular resolution of $ \leq 1^{\circ} $ and hence allow us to point back towards the source. As the main signature we focus on is the resultant muon, the interaction vertex is not required to lie inside the detector as in Ref.~\citep{HESEPaper,HESE3Year} and the effective volume is hence effectively enhanced. The results of an all-sky search, a search among a catalog of candidate neutrino emitters and stacked source catalog searches with a similar sample of events from the data collected between 2008-2011 are published in Ref.~\citep{IC79Paper}.  Here we update these analyses by adding the first year of data from the complete 86-string detector configuration, collected between May 2011 - May 2012. Five new stacking analyses based on newly available catalogs are also presented here.

In this paper we decribe the results of the first all-sky survey by IceCube looking for extended regions of neutrino emission. H.E.S.S. has surveyed the Galactic Plane looking for $\gamma$-ray emissions above 200\,GeV, revealing previously unknown extended regions emitting to\,TeV energies~\citep{HESSsurvey}. The Fermi/LAT survey above 100\,GeV also shows the same bright extended sources. These extended regions may be unidentified SNRs associated with molecular clouds, which are also expected to be spatially extended sources of neutrinos~\citep{MilagroNewHalzenPaper,MCNew}. Outside the Galaxy, large clusters of galaxies such as Virgo are promising neutrino emitters expected to have spatial extensions~\citep{2006PRD,galaxycluster,MuraseClusters,2008ApJClusters}.  It is therefore important not to limit the search for sources of neutrinos uniquely to point-like sources but also to extended regions as shown in Ref.~\citep{NeronovTeresaExtendedPaper}.

Section~\ref{sec3} describes the IceCube detector and the event selection for data from the first year of the completed detector. Event selections for data from the previous years of operation of the detector have been extensively described in Ref.~\citep{IC79Paper} and Ref.~\citep{IC40Paper}. The methodology used to combine data from different years and detector configurations and to optimize the searches for various source signal hypotheses is described in Sec.~\ref{sec4}. Section~\ref{sec5} presents the results of the analyses, which are discussed within the context of recent models of astrophysical neutrino emission.  Conclusions are drawn in Sec.~\ref{sec7}.

\section{Detector and Event Selection}
\label{sec3}

The IceCube Observatory is a cubic-kilometer-sized Cherenkov detector embedded in the ice at the geographic South Pole \citep{FirstYearPerformancePaper}.  Optimized to detect neutrinos above\,TeV energies, it consists of 5160 photomultiplier tubes (PMTs) instrumented along 86 cables (called strings) at depths of 1450 - 2450 \unit{m} beneath the surface of the ice sheet.  Each PMT is housed in a digital optical module (DOM), consisting of a pressure-resistant sphere with on-board digitization and calibration LEDs \citep{PMTPaper}.  The DOMs detect Cherenkov photons emitted by charged leptons that traverse the detector \citep{DOMMBPaper}.  This analysis uses data taken between April 2008 and May 2012.  During this period, IceCube ran in four different configurations. Three years of data are from the partial detector composed of 40-, 59-, and 79-strings, respectively, and are fully described in Ref.~\citep{IC79Paper}.  The following year of data was taken with the completed 86-string array. The used selection procedure and event reconstructions are similar to those applied to the previous data.

\subsection{Data Reduction and Reconstruction for the IC86-1 Data Sample}

Data acquisition is triggered by requiring four pairs of neighboring or next-to-neighboring DOMs to observe photoelectrons within a $\unit[5]{\mu s}$ time window. \unit[2.5]{kHz} of data satisfy this criterion. A combination of real-time filtering at the South Pole and subsequent offline CPU-intensive processing reduces the data rate to \unit[2]{Hz} by rejecting mis-reconstructed events.  At this stage the data are dominated by atmospheric muons from cosmic rays; both well reconstructed down-going muons in the southern sky and down-going muons mis-reconstructed as up-going muons in the northern sky.  The data is further reduced via quality cuts using simple reconstructions and event quality parameters followed by advanced likelihood-based muon reconstructions.  The simple reconstruction removes scattered photon hits before estimating the muon position and direction via a linear fit with reduced weights for outliers~\citep{ImprovedLineFitPaper}.  This fit serves as a seed for more advanced likelihood reconstructions, including the multi-photoelectron (MPE) likelihood.  This algorithm includes a probabilistic distribution function (PDF) that describes the scattering of photons in the ice, and is fully described in Ref.~\citep{AMANDAMuonRecoPaper}.

In the processing of data from the first year of the full detector, two new muon reconstructions were used to determine event directions and reject background. The first reconstructs the muon direction by applying the MPE likelihood four times.  Each iteration uses a bootstrapped pulse series, extracted randomly from the measured pulses.  This is done using a multinomial distribution weighted by charge, so that high charge pulses are more likely to be selected than low ones.  The results of these four reconstructions are averaged together to seed one reconstruction using the complete pulses.  Of these five fit results, the one with the best likelihood value is selected and saved.  Compared to the single-iteration MPE fit, this process reduces the rate of downgoing atmospheric muons mis-reconstructed as upgoing muons by 30\%, while improving the neutrino median angular resolution from $0.7^{\circ}$ to $0.6^{\circ}$ at \unit[30]{TeV}.

\begin{figure}[!th]
  \vspace{2mm}
  \centering
  \includegraphics[width=0.70\linewidth]{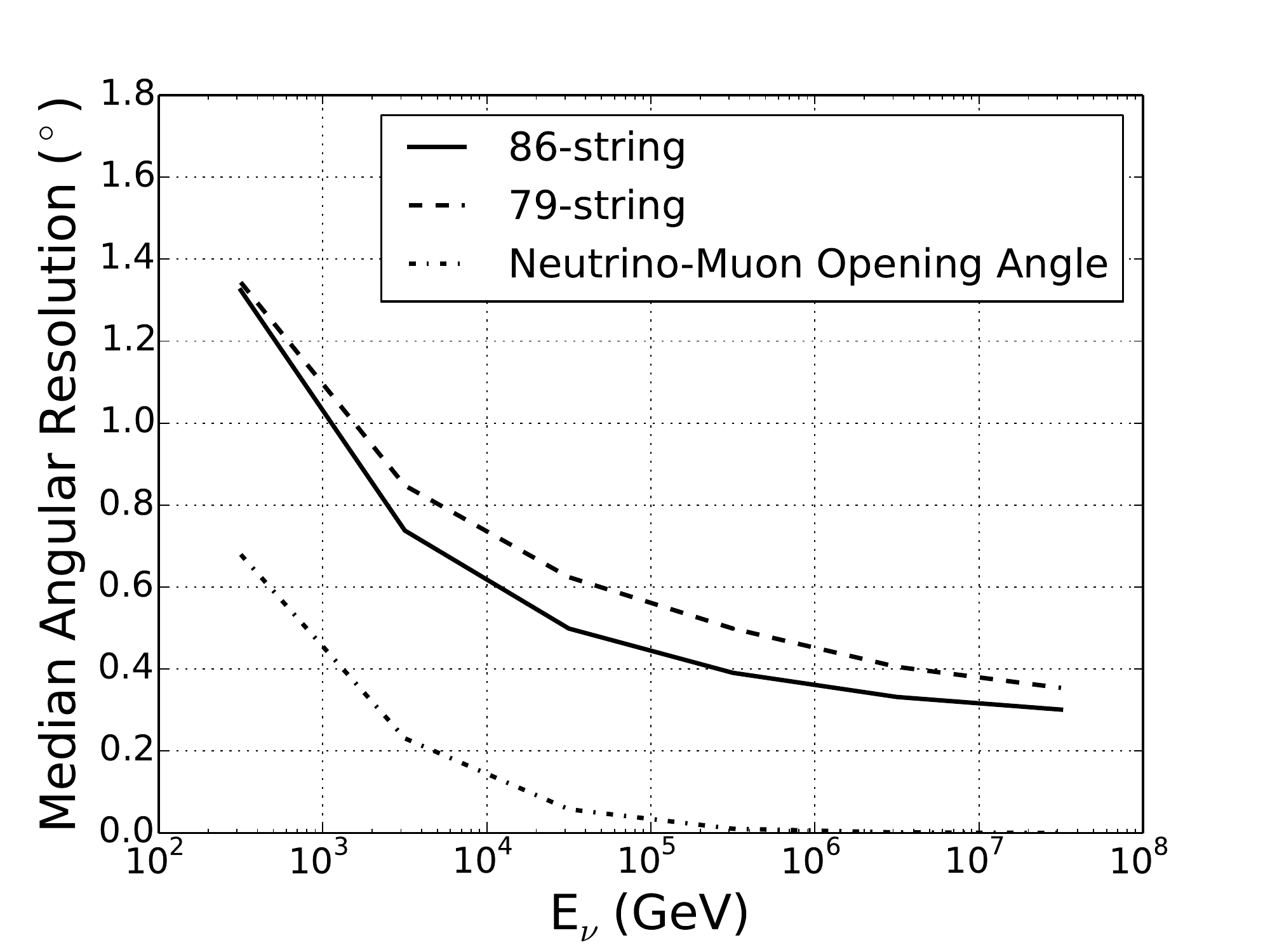}
    \caption{Median angular resolution (angle between reconstructed muon track and neutrino direction) as a function of neutrino energy for simulated northern hemisphere event samples from the 86-string (solid) and 79-string (dashed) detector configurations.  The improvement is due to the new reconstruction algorithm.  At 30\,TeV, the 40 and 59 string event selections (not shown) give angular resolutions of $\sim 0.8^{\circ}$ and $\sim 0.75^{\circ}$, respectively~\citep{IC79Paper}.  The dash-dotted line shows the median kinematic opening angle between the neutrino and muon.}
      \label{fig:angres}
\end{figure}

This iterative fit also serves as a seed for the second reconstruction algorithm, which provides a more accurate result by modeling the optical properties of the Antarctic ice sheet.  While previous reconstructions use analytic approximations to describe the timing distribution of Cherenkov photons arriving at a given PMT \citep{AMANDAMuonRecoPaper}, here we use a parametrization of a Monte Carlo simulation.  Photon transport is simulated using a depth-dependent model of scattering and absorption in the ice~\citep{SPICEPaper}.  The arrival time of a photon is a function of the orientation and depth of the muon source and the displacement vector between the muon and the receiving PMT.  Photons are simulated for different muon-receiver configurations, and a multi-dimensional spline surface is fit to the resulting arrival time distributions~\citep{PhotosplinePaper}.  These splines are used as PDFs in the MPE likelihood.  Compared to previous IceCube point source analyses~\citep{IC79Paper}, this reconstruction algorithm leads to a 26\% improvement in neutrino median angular resolution at 30\,TeV (see Fig.~\ref{fig:angres}).  As carried out in previous years, the uncertainty in the angular reconstruction for each event is estimated by fitting a paraboloid to the likelihood space around the reconstructed direction, following the method described in \citep{IC40Paper, ParaboloidPaper}.

After reconstructing the direction of each event, a separate algorithm fits for the muon energy loss along its track.  In the fourth year of data, the energy reconstruction uses an analytic approximation to model the muon light yield at the receiving DOMs as a function of the orientation and depth of the muon~\citep{EnergyRecoPaper}.

\subsection{Selection of the Final Sample}

From the \unit[2]{Hz} of remaining data (still dominated by the atmospheric muon background), \unit[4.8]{mHz} of events are selected for the final analysis sample.  In the northern sky the mis-reconstructed muon background can be mostly eradicated to isolate a nearly pure sample of up-going atmospheric neutrinos.  This is done using a classification algorithm, Boosted Decision Trees (BDTs).  Similar to previous IceCube point source analyses~\citep{IC79Paper}, we trained four BDTs in two zenith bands to separate astrophysical neutrino signal from the atmospheric muon background.  Cuts on the BDT output scores are optimized to achieve the best discovery potential for both E$^{-2}$ and E$^{-2.7}$ signal spectra.  This event selection covers the entire Northern Hemisphere and extends $5^{\circ}$ above the horizon, where the Earth and glacial ice still provide a shield from the cosmic ray background.

At an angle of more than $5^{\circ}$ above the horizon, a pure neutrino sample cannot be isolated from the high-energy atmospheric muon bundles, which are multiple muons from the same air shower that mimic neutrinos. The background can be reduced by introducing quality cuts and using parameters that select neutrinos and reject muon bundles.  One BDT is trained for the entire region using data to describe the background and an E$^{-2}$ neutrino simulation for signal.  Of the eleven variables used in training the BDT, three exploit differences between single muons and bundles.  These parameters rely on event topology and energy loss information.  Large muon bundles consist of many low-energy muons that typically lose energy at a constant rate as they traverse the detector. Photons from these muon bundles are detected within a wider time range.  High-energy neutrino-induced muons instead have relatively stochastic energy loss profiles and narrower photon timing distributions.  Likelihood ratios are constructed to judge whether a given data event has timing and energy loss properties more consistent with the simulated signal or the estimated background, and are included in the BDT.  To obtain the final sample, a cut on the BDT score is varied with zenith to account for the zenith-dependent properties of the background.

\begin{figure}
\includegraphics[width=0.5\linewidth]{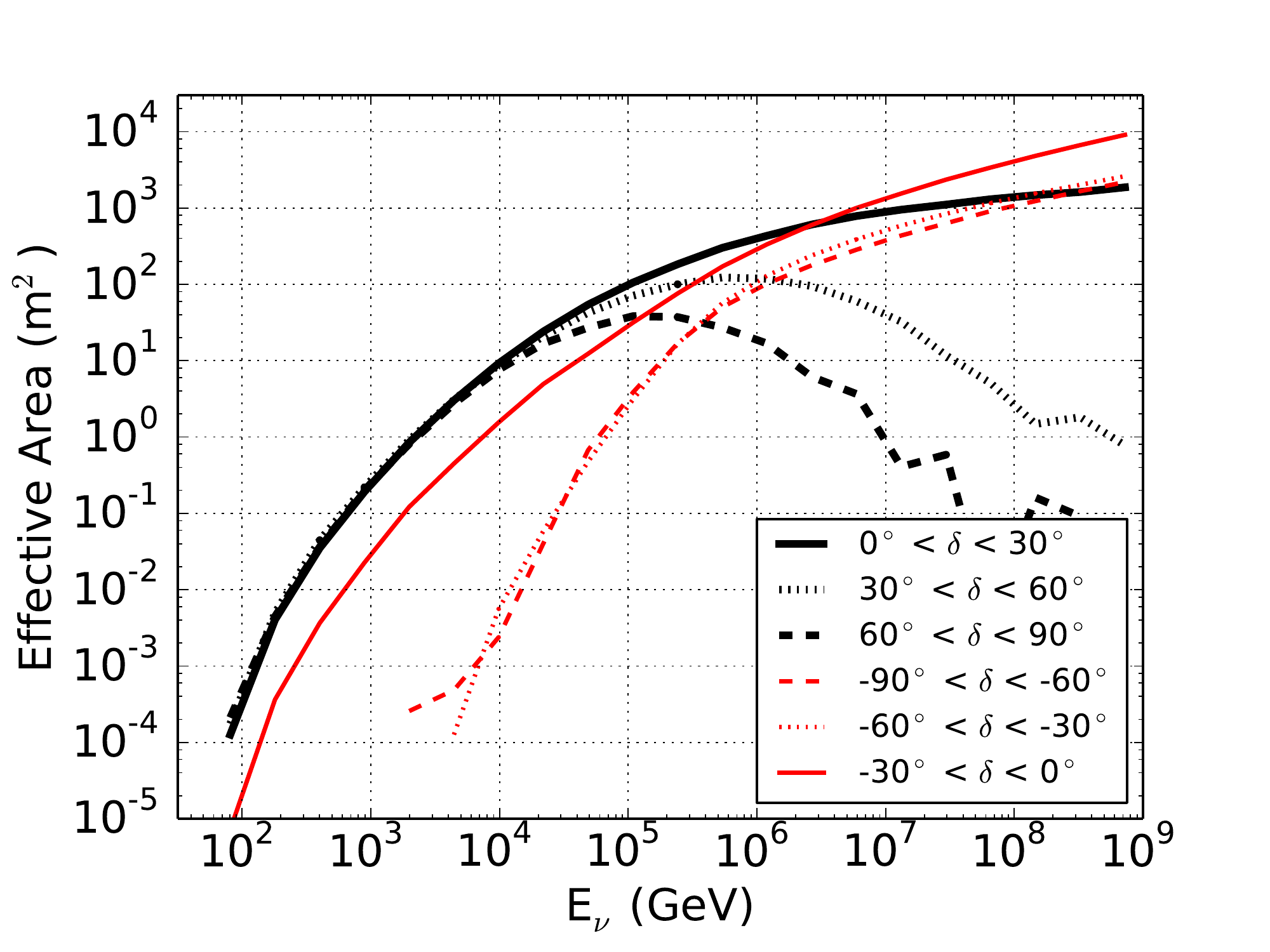}
\includegraphics[width=0.5\linewidth]{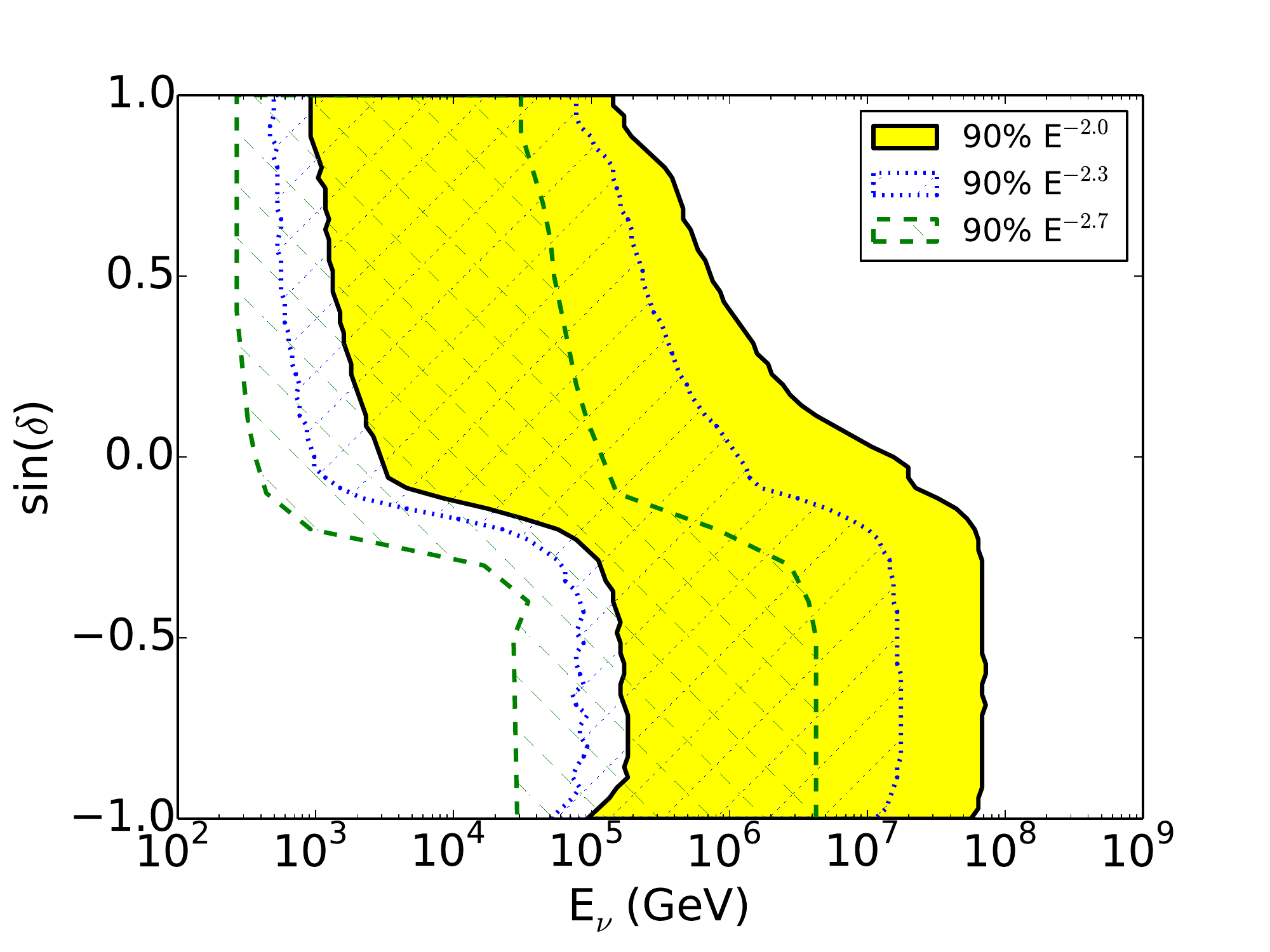}
\caption{Left: Neutrino effective area for the 86-string detector as a function of primary neutrino energy for six declination bands.  The effective area is the average of the area for $\nu_{\mu}$ and $\bar{\nu_{\mu}}$. Right: Central 90\% energy region for simulated neutrino events as a function of declination.  This defines the region where the upper limits for $E^{-2}$, $E^{-2.3}$, and $E^{-2.7}$ source spectra are valid.}
\label{fig:EffArea}
\end{figure}

\begin{figure}
\centering
\includegraphics[width=4.in]{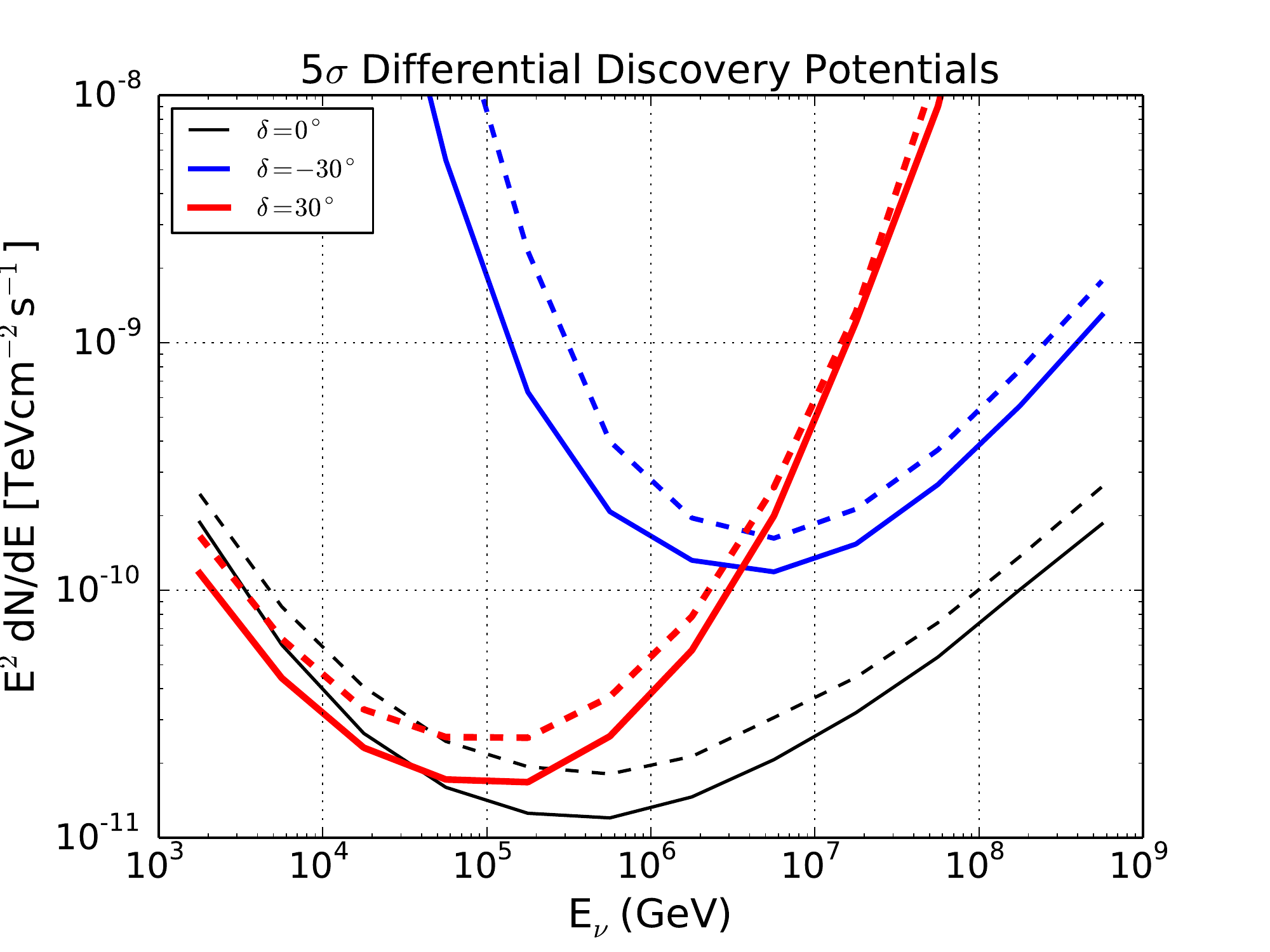}
\caption{Discovery flux as a function of the neutrino energy at 5$\sigma$ confidence level, for three different declinations (solid lines).  Point sources with an $E^{-2}$ spectrum are simulated over a half-decade in energy, and the flux in each bin required for discovery forms the curve above.  Results from the previous analysis with 3 years of the data are shown with dashed lines.}
  \label{fig:discoE2}
\end{figure}

The final data sample for the first year of operation of the 86-string detector has 138,322 events, of which approximately half are in the northern hemisphere.  The livetime and rates for all four years of detector data are summarized in Table~\ref{tab:lifetimes}.  The neutrino effective area for this selection and the central 90\% energy region for three signal spectra are shown in Figure~\ref{fig:EffArea}.  The effective area reaches it's maximum near the horizon.  Far below the horizon high-energy neutrinos suffer from absorption in the Earth.  Above the horizon the cuts necessary to remove the background remove a significant portion of the lower-energy signal.  As a result the analysis is sensitive to the widest neutrino energy range near the horizon, while in the southern hemisphere the sensitivity rapidly deteriorates at lower energies.  The discovery potential as a function of energy and declination is shown in Fig.~\ref{fig:discoE2}.  Compared to the 3-year point source analysis~\citep{IC79Paper}, the addition of the first year of data from the completed detector including improved reconstruction and background rejection techniques leads to a 40 $-$ 50\% improvement in the discovery potential, with larger gains at energies below \unit[1]{PeV} in the southern hemisphere.

\begin{table*}
\begin{center}
  \caption{Summary for four different IceCube configurations for point
    source analyses: The expected atmospheric neutrino rate from MC
    simulation weighted for the model in Ref.~\citep{Honda} and numbers of
    up- and down-going events at final selection level.  The upgoing data are dominated by atmospheric neutrinos, while data in the downgoing region are dominated by atmospheric muons.}
\vspace{0.2cm}
\small{
\begin{tabular}{ccccc}
\toprule
no. of strings & live-time [days] & atm. $\nu$s & \# up-going &  \# down-going\\
\midrule 
40  &  376 &  40/day & 14,121 & 22,779 \\
59  &  348 &  120/day& 43,339 & 64,230 \\
79  &  316 &   180/day& 50,857 & 59,009 \\
86  &  333 &   210/day& 69,227 & 69,095 \\
\bottomrule
\end{tabular}}
\label{tab:lifetimes}
\end{center}
\end{table*}

\section{The Likelihood Search Method}
\label{sec4}

Point-like sources of neutrinos in the sky can be identified by searching for clusters of events significantly incompatible with the atmospheric muon and neutrino background. The significance is estimated by using an unbinned maximum likelihood ratio test as described in Ref.~\citep{method}. The method is expanded to allow for the combination of data from different detector geometries as described in Ref.~\citep{IC79Paper}. In addition to spatial clustering, this method also uses the energies of the events to identify signal events which are expected to have a harder spectrum than that of atmospheric neutrinos and muons. The energy response expected from a neutrino signal from a point source in the sky is modeled using simulation. Since the final event selections are still background dominated, the background estimate is done using real data.

In time integrated searches for a point-like source, the signal PDF $\mathcal{S}^{j}_i$ for event $i$ observed in detector geometry $j$ is given by:

\begin{equation}
\label{eq:signalpdf}
\mathcal{S}^j_i = S^{j}_{i} (|\vec{x}_i-\vec{x}_{s}|,\sigma_{i}) \mathcal{E}^{j}_{i} (E_i, \delta_i, \gamma)
\end{equation}

Here, the spatial contribution to the PDF is given by $S^{j}_{i}$, which depends on the angular uncertainty of the event $\sigma_i$, and the angular difference between the reconstructed direction of the event and the direction of the source. This probability is modeled as a 2-dimensional Gaussian:

\begin{equation}
\label{eq:space}
S^j_i = \frac{1}{2\pi\sigma^2_i}e^{-\frac{|\vec{x}_i-\vec{x}_{s}|^2}{2\sigma^2_{i}}}.
\end{equation}

The contribution from energy $\mathcal{E}^{j}_{i} (E_i, \delta_i, \gamma)$ is described in Ref.~\citep{method}.

\begin{figure}[!th]
  \vspace{5mm}
  \centering
  \includegraphics[width=0.70\linewidth]{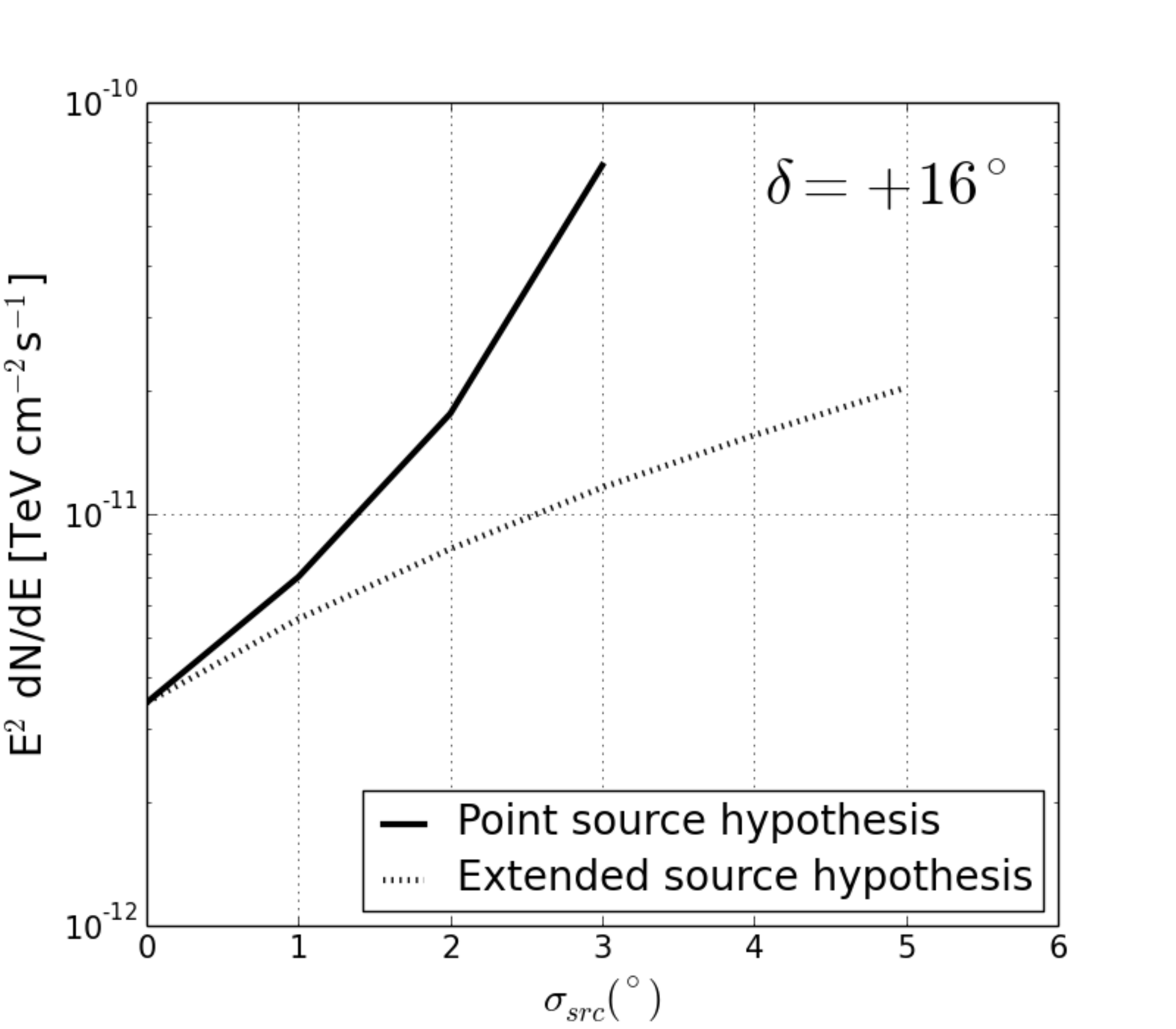}
  \caption{Flux needed for a 5$\sigma$ discovery from a hypothetical source at $\delta=16^{\circ}$ as a function of the source extension for the point source signal hypothesis (solid line) and the extended signal hypothesis with the correct extension (dotted line).}
  \label{fig:Extendedmotivation}
 \end{figure}

When searching for spatially extended sources the value of $\sigma_i$ is replaced with $\sigma_{i}^{\text{eff}} = \sqrt{\sigma_i^2 + \sigma_{\text{src}}^2}$ where $\sigma_{\text{src}}$ is the width of the source. Fig.~\ref{fig:Extendedmotivation} shows the flux needed for a 5$\sigma$ discovery for a source located at a given declination as a function of the source extension. The results for two different signal hypotheses are shown; in one the source is always assumed to have no extension while in the other the correct source extension is included in the likelihood description. Naturally, for sources that are truly extended the extended hypothesis is more powerful than the point source assumption. As the real extension of the source increases, the analysis method which assumes that the source is point-like performs worse than the one that takes the extension of the source in to account.

To further enhance discovery potentials and sensitivity, stacked searches can be carried out for specific catalogs of similar candidate neutrino sources.

The following is a description of all the searches performed with the four
years of IceCube data (similar to those performed in Ref.~\citep{IC79Paper}):

\subsection{{\bf All-Sky Searches}}
These searches are carried out to look for evidence of a source anywhere in the sky and are not motivated by any prior information regarding the position of the sources. The likelihood is evaluated in each direction in the sky. In these searches the number of effective trials is very high and is related to the angular resolution of the telescope and the source extension hypotheses. In order to correct for the trial factor, the same experiment is repeated on an ensemble of scrambled data and the probability of observing a more significant spot than the one observed is obtained.

\begin{description}
 \item [All Sky Point Source Scan] The all-sky scan for point sources of neutrinos that has previously been carried out on data from the incomplete detector configurations is updated to include the first year of data from the complete 86 string detector. In this search the likelihood is evaluated in steps of $0.1^{\circ} \times 0.1^{\circ}$ within the declination range -85$^{\circ}$ to +85$^{\circ}$ beyond which the scrambling technique is no longer effective.
 
 \item [All Sky Extended Source Scans]  The search for extended sources is performed in a similar fashion to the all-sky point source searches. In this case the sky is divided into a grid of $0.5^{\circ} \times 0.5^{\circ}$ in a similar declination range. For this search a source extension needs to be assumed for the signal. We carry out five different all-sky scans assuming extensions in step of one degree, from 1$^{\circ}$ to 5$^{\circ}$. An additional trial factor needs to be considered from the additional number of sky scans, however this factor can be conservatively assumed to be 5. 
\end{description}

\subsection{{\bf Searches Among List of 44 Candidate Sources}}
\label{subsec41}
  
In order to reduce the large number of effective trials associated with scanning the entire sky, we also performed a search for the most significant of 44 {\it a priori} selected source candidates. The sources in this list have been selected according to observations in $\gamma$-rays or astrophysical models predicting neutrino emission.

\subsection{{\bf Stacking Searches}}

Several sources of the same type may emit fluxes that are individually below the discovery potential but detectable as a class when summed up using the stacking technique. Here we report on the different catalogs of sources that have similar spectral behavior based on $\gamma$-ray observations or astrophysical models predicting neutrino emission. For these searches, the signal PDF $\mathcal{S}^{j}_i$ of Eq.~\ref{eq:signalpdf} is modified to accommodate multiple sources (see Ref.~\citep{IC40Paper}). A prior knowledge of the expected luminosities of these sources can be utilized to weight the contribution of each source in the total signal PDF to make the search optimal for that signal hypothesis. Alternatively, an equal-weighting can be applied if there is no preferred model. In the following section we summarize all the stacking searches performed with 4 years of data. Most of these searches are updates from the previous results using 3 years of data~\citep{IC79Paper}.

{\bf \underline{Updated searches}} : These searches have been previously carried out on three years of data~\citep{IC79Paper} and are now updated to include data from the first year of operation of the completed 86-string detector. 

\begin{description}

\item[6 Milagro TeV gamma-ray sources.] The authors of the model that motivated the original analysis have hence updated the models to reflect the newer $\gamma$-ray observations ~\citep{MilagroNewHalzenPaper}. For this reason, in this search an equal weight is used for each source in the likelihood with the intention of keeping our sensitivity optimal for all possible signal hypothesis.
\item[127 local starburst galaxies.] Sources compiled in Table A.1 in Ref.~\citep{Starburst}.
\item[5 nearby clusters of galaxies.] This search tests four models assuming
  different CR spatial distribution within the
  source~\citep{galaxycluster}. 

  \item[10 SNRs associated with molecular clouds.]  This search is now updated to include more sources in the southern sky owing to our increased sensitivity in the southern sky due to new background rejection techniques. From the exhaustive online catalog SNRCat~\citep{SNRCat}, we select sources with confirmed molecular clouds associations. In order to keep the most promising neutrino emitters within the catalog, only sources that have been observed in the TeV or are younger than 10,000 yrs (potentially in the Sedov Blast wave phase~\citep{Sedov} of expansion) are considered. 
  The catalog contains 4 SNRs associated with molecular clouds in the northern sky \citep{milagro2007,W51CFermi,W51CMilagro,W44Fermi,FermiSNR,W51CHess} that were previously considered in Ref.~\citep{IC79Paper}, and 6 newly introduced sources from SNRCat in the southern sky. These 6 sources are Sgr A East, Kes 75, 3C391, RX J1713.7-3946, CTB 37A and 1FGL J1717.9-3729.

\item[233 Galaxies with super-massive black holes.]  A sample of AGNs within the GZK ~\citep{GZK} radius as cataloged by Ref.~\citep{bh} keeping only sources more massive than $5 \times 10^{8}$ solar masses.

\end{description}

 {\bf \underline{New searches:}} These are new searches introduced with the inclusion of the first year of data from the completed 86-string detector.

\begin{description}

\item[10 Galactic Pulsar Wind Nebulae.]
 Pulsar Wind Nebulae (PWN) are potential emitters of neutrinos~\citep{PWNPaper}. We carry out a stacked search for neutrinos coming from known PWNs within the Galaxy. From the confirmed PWNs in SNRCat ~\citep{SNRCat}, we look at sources that are younger than 10,000 years as only younger PWNs are efficient accelerators~\citep{PWNPaper}. We leave out sources that are already considered by the search for SNRs associated with molecular clouds. These criteria are fulfilled by 3 sources in the northern sky, namely the Crab Nebula, DA 530, G054.1+00.3 and 7 sources in the southern sky including the Pencil Nebula, W33 and MSH 11-54. These sources are weighted in likelihood by the inverse of their median age as provided by SNRCat ~\citep{SNRCat} to account for the higher fluxes expected from the youngest PWNs ~\citep{PWNPaper}.
 
\item[30 Galactic SNRs.]
  Galactic SNRs~\citep{SNRCat} which neither have confirmed molecular cloud associations nor are PWNs are considered in this stacking search. As in the searches for PWNs and SNRs with Molecular Cloud associations, a cut on the SNR age is applied and only those younger than 10,000 years are selected~\citep{Castro}.  This requirement is met by 30 sources in total where 20 are located in the southern sky and 10 in the north. The inverse of the median age as provided by SNRCat ~\citep{SNRCat} is used as the weight for each source in likelihood in order to account for the fact that we expect the highest fluxes to come from the youngest SNRs. Remnants of recent prominent Supernovae such as Casseopeia A and Tycho are considered within this search.
  
\item[Blazars catalogs]
  Three Blazar catalogs were composed from the Fermi LAT Second AGN Catalog~\citep{FermiAGNCat} to allow for optimized analyses of the corresponding object classes. 
  The first catalog contains Flat Spectrum Radio Quasars (FSRQ) which as suggested by their broad line regions are thought to provide efficient photomeson production~\citep{DermerModelPaper} in dense soft photon targets.
  The second set is formed by low-frequency peaked (LSP) BL Lac objects that are predicted to show a significant contribution from pion decays to the overall gamma-emission in the Synchrotron Proton Blazar Model~\citep{MueckeModelPaper}. 
  Finally, p-p interaction models are covered by a catalog of the BL Lac objects with particularly hard gamma spectra and correspondingly large effective areas for neutrinos in IceCube~\citep{NeronovModelPaper}.

  The source selection and weighting for the FSRQ and LSP BL Lac catalogs, assuming prevalence of photo-hadronic neutrino production is based on the Fermi LAT gamma-flux. This motivates a weighting that is based on the measured gamma-fluxes but assumes the same spectral index for all sources (hereby denoted by W1). 

In proton-proton interaction models, the energy spectrum of the produced neutral secondaries follows the initial cosmic ray spectrum down to a threshold below \unit[1]{GeV}. The observation of the gamma-spectrum thus allows for a direct prediction of the proton spectrum behavior in the TeV range, which can be extrapolated to PeV energies to estimate the neutrino spectrum. Such an approach is not as easily possible for proton-gamma interaction models, as these typically have a lower energy threshold above TeV energies so that the photon (and neutrino) spectrum below the threshold does not allow for the derivation of the proton spectrum \citep{NeronovModelPaper}.
Hence, the third catalog of hard $\gamma$-spectrum BL Lac objects motivates a selection and weighting based on the number of detectable neutrinos derived from the spectral shape measured by Fermi LAT (hereby denoted by W2).
	
Due to the variety of Blazar models and the large model uncertainties, both weighting schemes are applied to all three catalogs. Sources with negligible weights in both weighting schemes are discarded, resulting in 33 FSRQs, 27 LSP BL Lac objects and 37 hard $\gamma$-spectrum BL Lac objects.

  This stacked search for blazars uses a reprocessed data set of the 79 string configuration that incorporates the new reconstruction methods presented in this work for IC-86, which were not yet available at the time of the previous analyses.
\end{description}

\section{Results and Implications}
\label{sec5}

In this section we summarize all the results from the different searches and their implication on astrophysical models of neutrino emission.  While no significant excess has been found in any of the searches and all results are consistent with the background-only hypothesis, this has allowed us to set upper limits that exclude some of the models.
\subsection{{\bf All-Sky Searches}}
\subsubsection{All-Sky Point Source Scan}

\begin{figure}[t!]
\hspace{-2cm}
\includegraphics[width=1.25\textwidth]{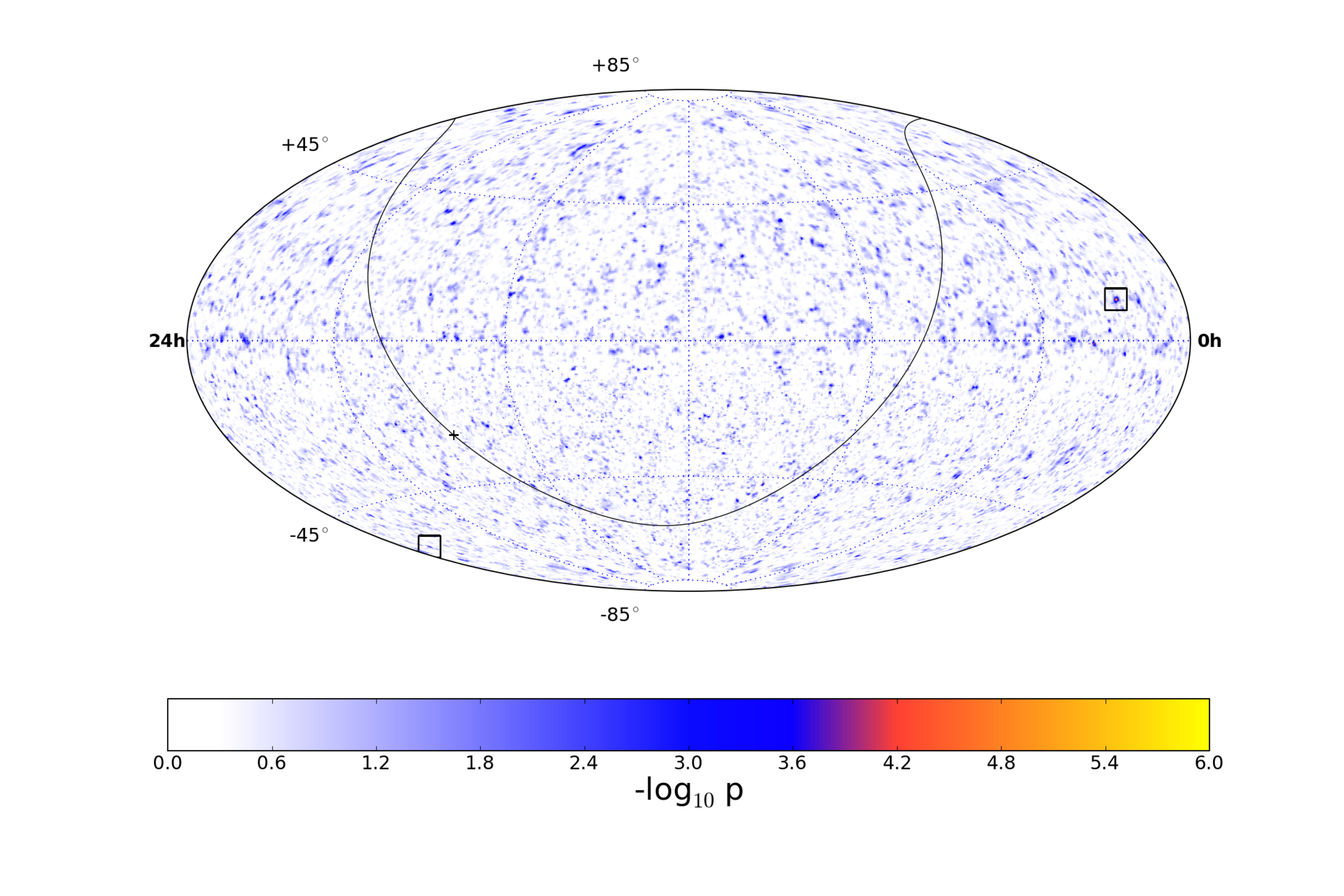}
\vspace{-1cm}
\caption{Pre-trial significance skymap in equatorial coordinates (J2000) of the all-sky point source scan for the combined four year data sample. The black line indicates the Galactic plane, and the black plus sign indicates the Galactic Center. The most significant fluctuation in each hemisphere is indicated with a square marker.}
\label{fig:skymap}
\end{figure}

Figure \ref{fig:skymap} shows the result of the all-sky scan for point sources in terms of significance at each location in the sky given in equatorial coordinates. The most significant deviation in the northern sky has a pre-trial p-value of $4.81 \times 10^{-6}$, and is located at $29.25^{\circ}$ r.a. and $10.55^{\circ}$ dec.  At this location, the best fit values of the number of source events, $\hat{n}_s$, and signal spectral index, $\hat{\gamma}$, are 43.0 and 2.88, respectively.  In the southern sky, the most significant deviation has a pre-trial p-value of $6.81 \times 10^{-6}$ and is located at $347.95^{\circ}$ r.a. and $-57.75^{\circ}$ dec.  Here, the best fit values of $\hat{n}_s$ and $\hat{\gamma}$ are 13.0 and 3.95, respectively.  After accounting for the trial factor associated with scanning the sky for the most significant spots, the post-trial p-values are 0.23 for the spot located in the northern sky and 0.44 for the spot located in the southern sky.

\subsubsection{All-Sky Scans for Extended Sources}

\begin{figure}[!th]
  \vspace{5mm}
  \centering
  \includegraphics[width=.85\textwidth]{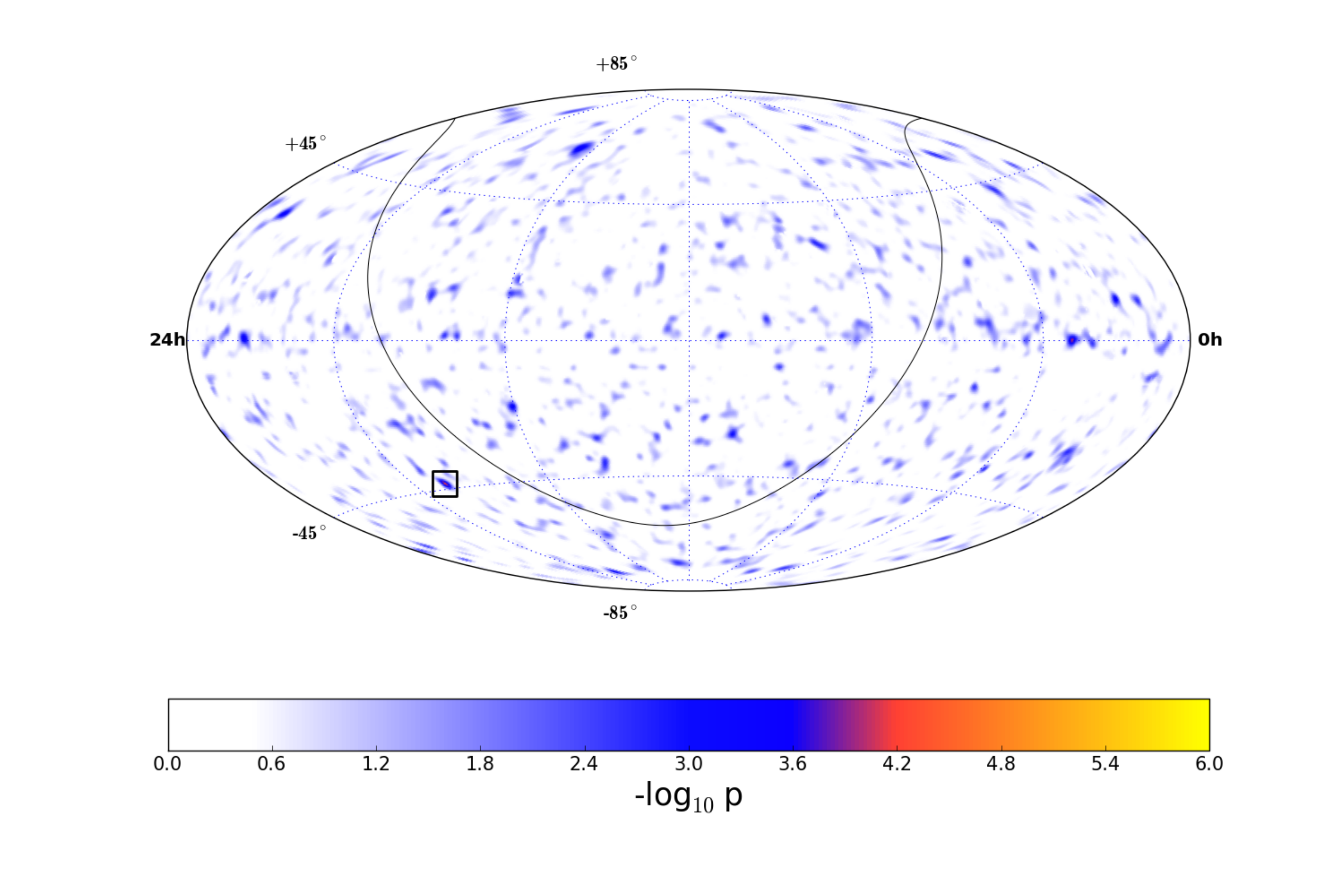}
  \caption{Pre-trial significance skymap from the all-sky scan for sources of $1^{\circ}$ extension in equatorial coordinates. The black line indicates the Galactic Plane. The most significant fluctuation is indicated with a square marker. }
  \label{fig:ExtSkyMap1}
 \end{figure}

\begin{figure}[!th]
  \vspace{5mm}
  \centering
  \includegraphics[width=.87\textwidth]{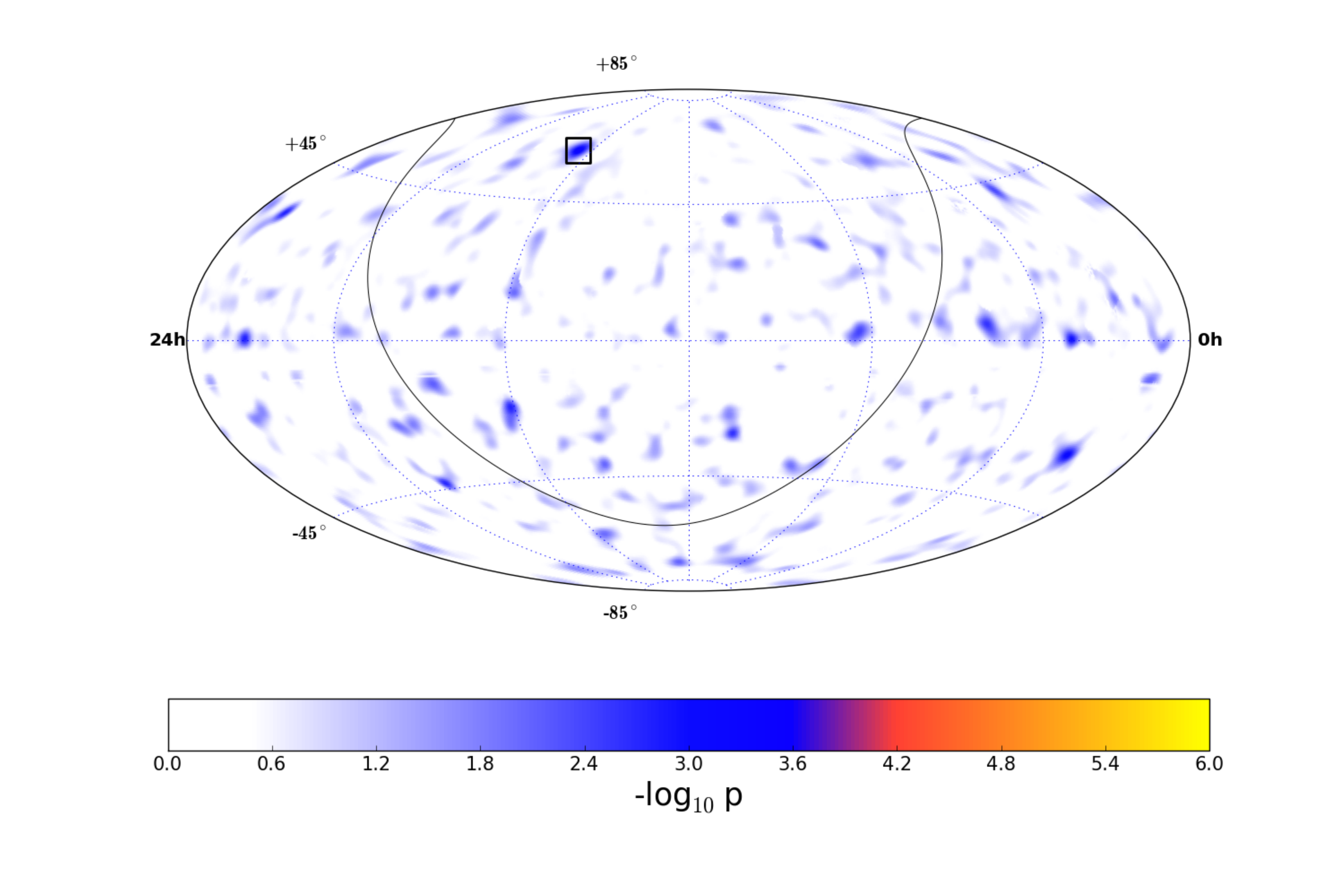}
  \caption{Same as Fig.~\ref{fig:ExtSkyMap1} but for sources of $2^{\circ}$ extension. }
  \label{fig:ExtSkyMap2}
 \end{figure}

\begin{figure}[!th]
  \vspace{5mm}
  \centering
  \includegraphics[width=.87\textwidth]{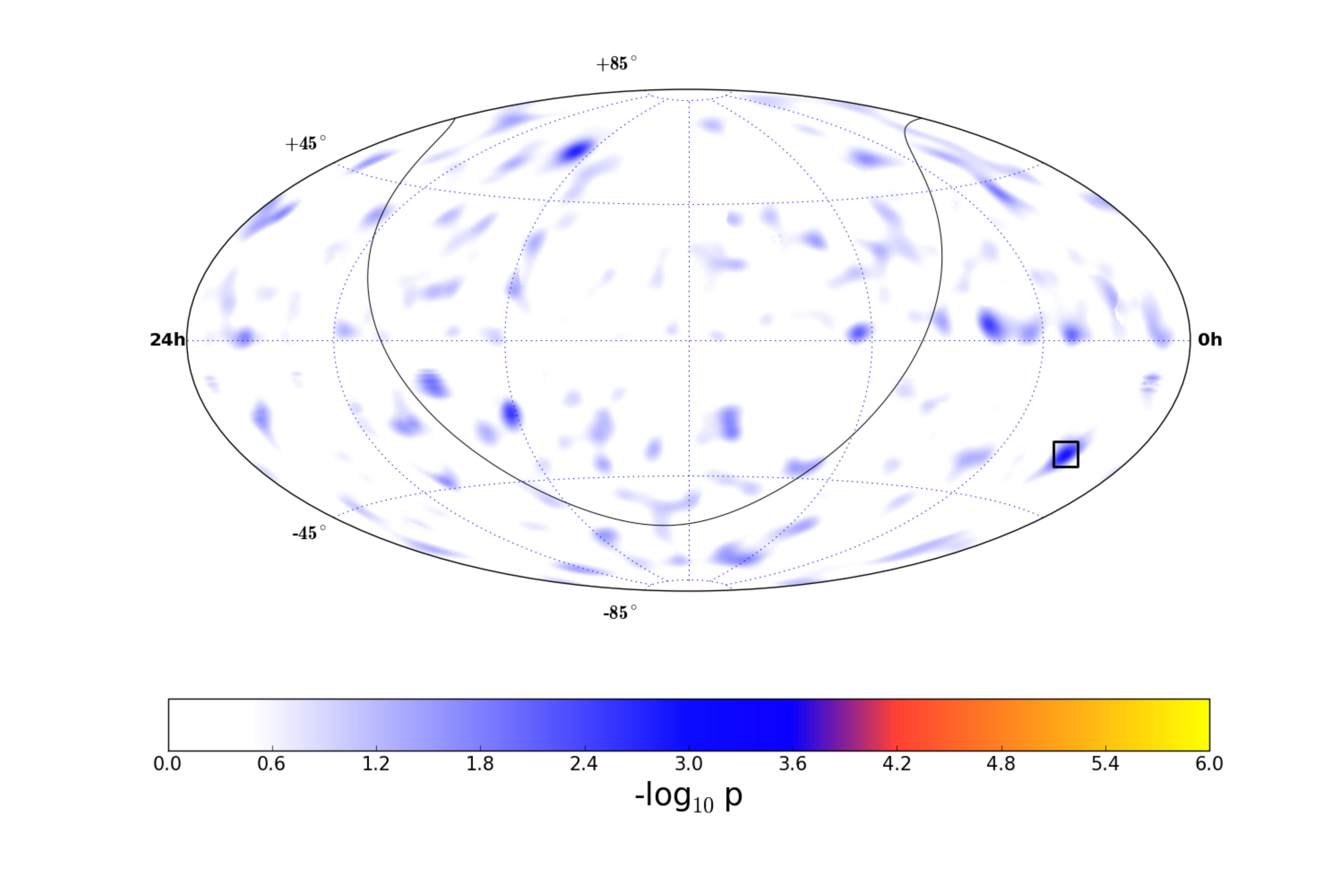}
  \caption{Same as Fig.~\ref{fig:ExtSkyMap1} but for sources of $3^{\circ}$ extension.}
  \label{fig:ExtSkyMap3}
 \end{figure}

\begin{figure}[!th]
  \vspace{5mm}
  \centering
  \includegraphics[width=.87\textwidth]{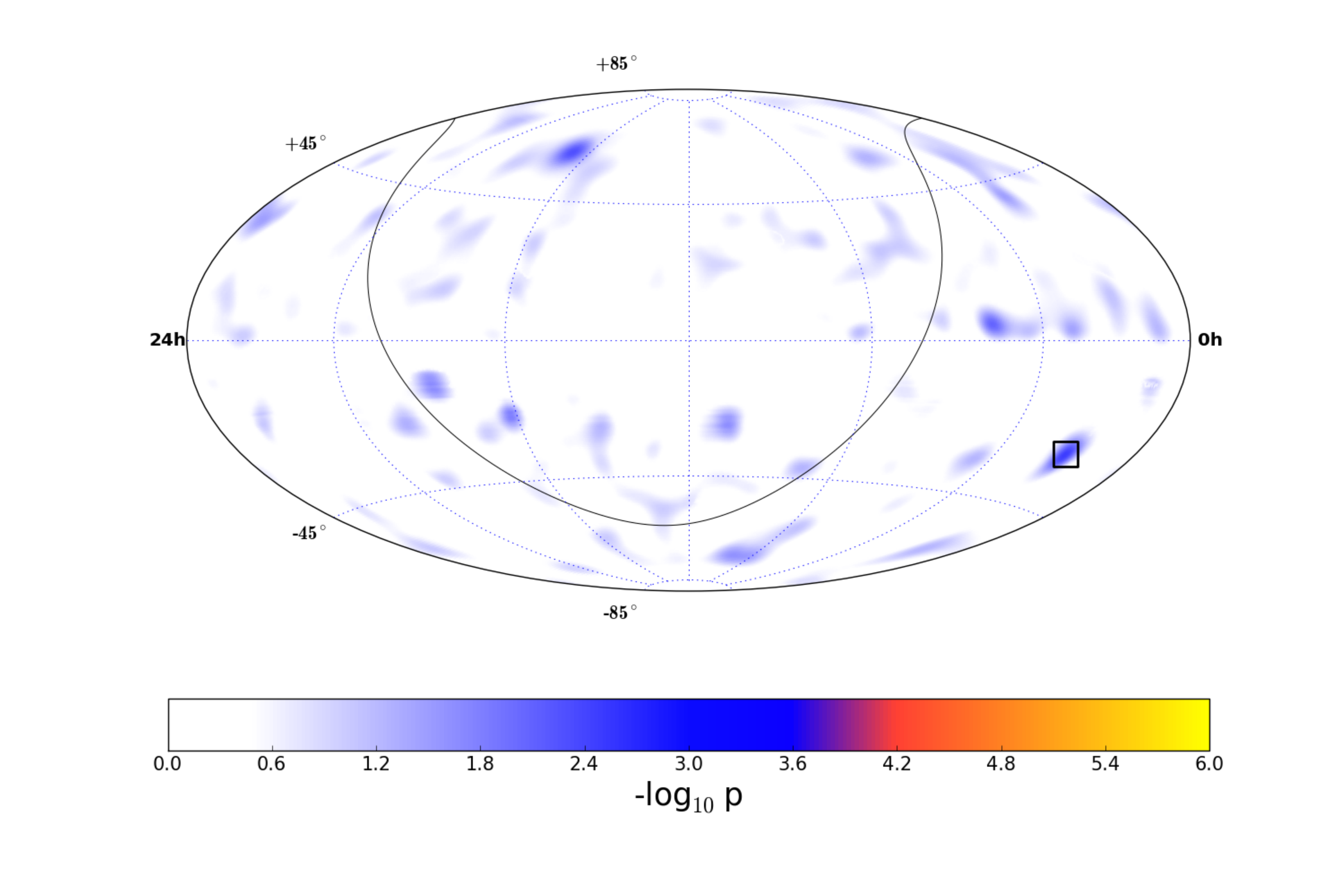}
  \caption{Same as Fig.~\ref{fig:ExtSkyMap1} but for sources of $4^{\circ}$ extension.}
  \label{fig:ExtSkyMap4}
 \end{figure}

\begin{figure}[!th]
  \vspace{5mm}
  \centering
  \includegraphics[width=.87\textwidth]{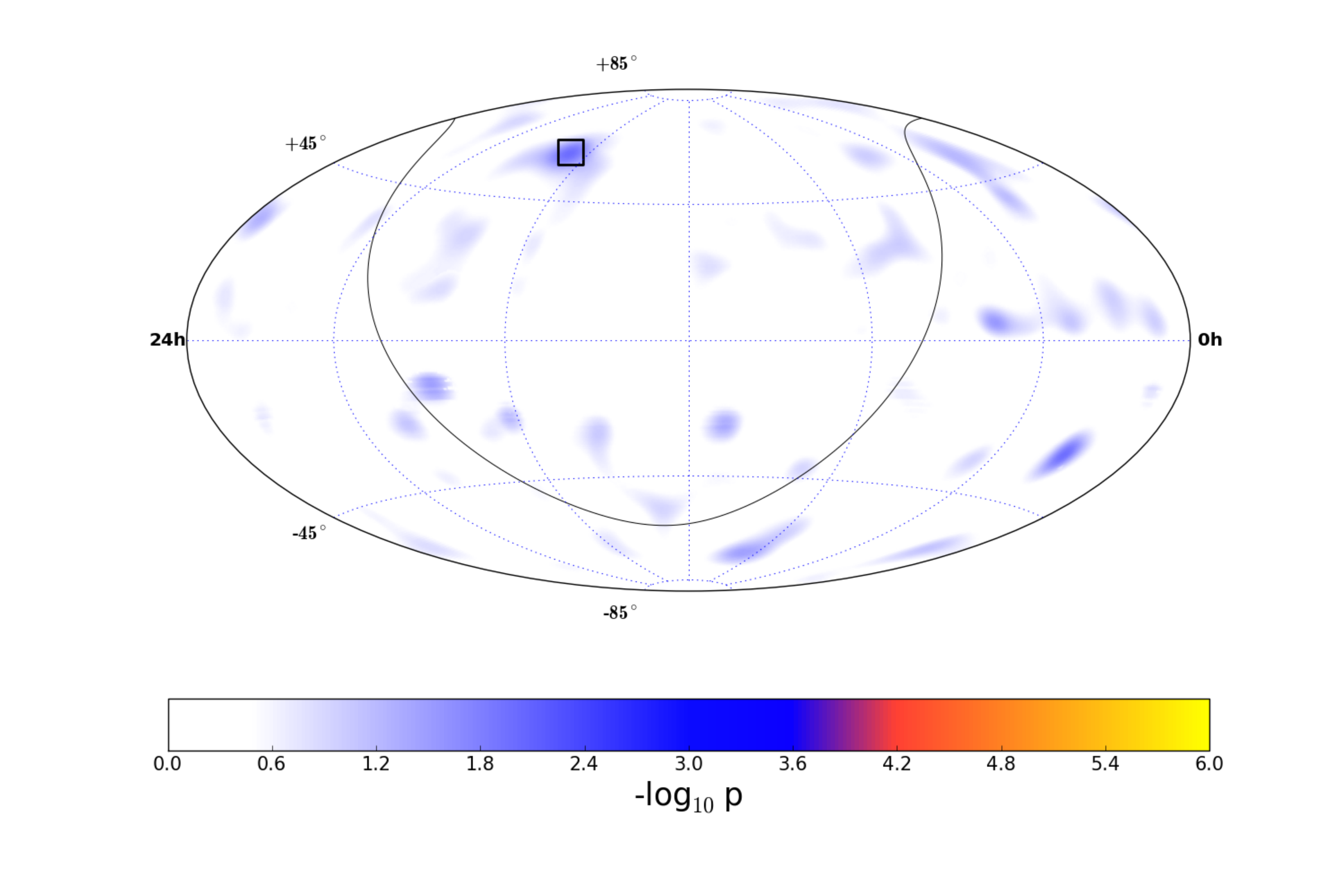}
  \caption{Same as Fig.~\ref{fig:ExtSkyMap1} but for sources of $5^{\circ}$ extension.}
  \label{fig:ExtSkyMap5}
 \end{figure}

Table ~\ref{tab:EXTable} summarizes the most significant hotspots in the sky from the scans for sources of various extensions. All observations were compatible with the background hypothesis. Figures \ref{fig:ExtSkyMap1} to \ref{fig:ExtSkyMap5} show the corresponding skymaps for $1^{\circ},\,2^{\circ}, \,3^{\circ}, \,4^{\circ}$ and $5^{\circ}$
extension respectively.

\clearpage
\begin{table}[htbp]
 \centering
 \begin{tabular}[b]{l | c  | c| c |c |c |c}
  \toprule
  Extension [$^{\circ}$] & r.a. [$^{\circ}$] & dec. [$^{\circ}$]  & $\hat{n}_{S}$ & $\hat{\gamma}$ & p-value (pre-trial) & p-value (post-trial)\\
  \midrule[\heavyrulewidth]
  1$^{\circ}$ & 286.25 & -43.25 & 49.6 & 2.65 & $6.75 \times 10^{-5}$ & 0.58\\
  2$^{\circ}$ & 248.75 & 62.75 & 58.2 & 2.38 & $5.52 \times 10^{-4}$ & 0.87\\
  3$^{\circ}$ & 30.75 & -30.25 & 93.6 & 3.10 & $1.22 \times 10^{-3}$ & 0.81\\
  4$^{\circ}$ & 30.75 & -30.25 & 99 & 3.10 & $3.29 \times 10^{-3}$ & 0.81\\
  5$^{\circ}$ & 251.75 & 61.25 & 102 & 2.54 & $1.06 \times 10^{-2}$ & 0.91\\
  \bottomrule
 \end{tabular}
\caption{Summary of the results from the extended all-sky survey. The coordinates of the most significant spots located for each source extension hypothesis are given together with the respective {\it p}-values.}
\label{tab:EXTable}
\end{table}

\clearpage

Since filtering streams, reconstructions and detector configurations evolved with time, we also examined each of the four years of data independently as an a posteriori cross-check. The largest fluctuation was observed for the one degree extension hypothesis in data from the 79 string configuration at 266.75 r.a. and 13.25 dec, where 0.35\% of scrambled maps in that year resulted in a fluctuation more significant than the one observed. Since we scanned over 5 different extensions for every year, the corresponding trial-corrected p-value is ~7.2\%, well compatible with a background fluctuation. The
hot-spot seems to be driven by a single well-reconstructed very high-energy event which, when folded with the wider source template, overlaps with some nearby lower energy ones. From calibration using the shadow of the moon \citep{moonpaper}, there is no evidence for a systematic error in IceCube's point spread function that could lead to the observed spread for events originating from a point-like source. The region is not significant in any of the other years of data. 

\subsection{List of 44 Candidate Sources}

\begin{figure}
\includegraphics[width=\linewidth]{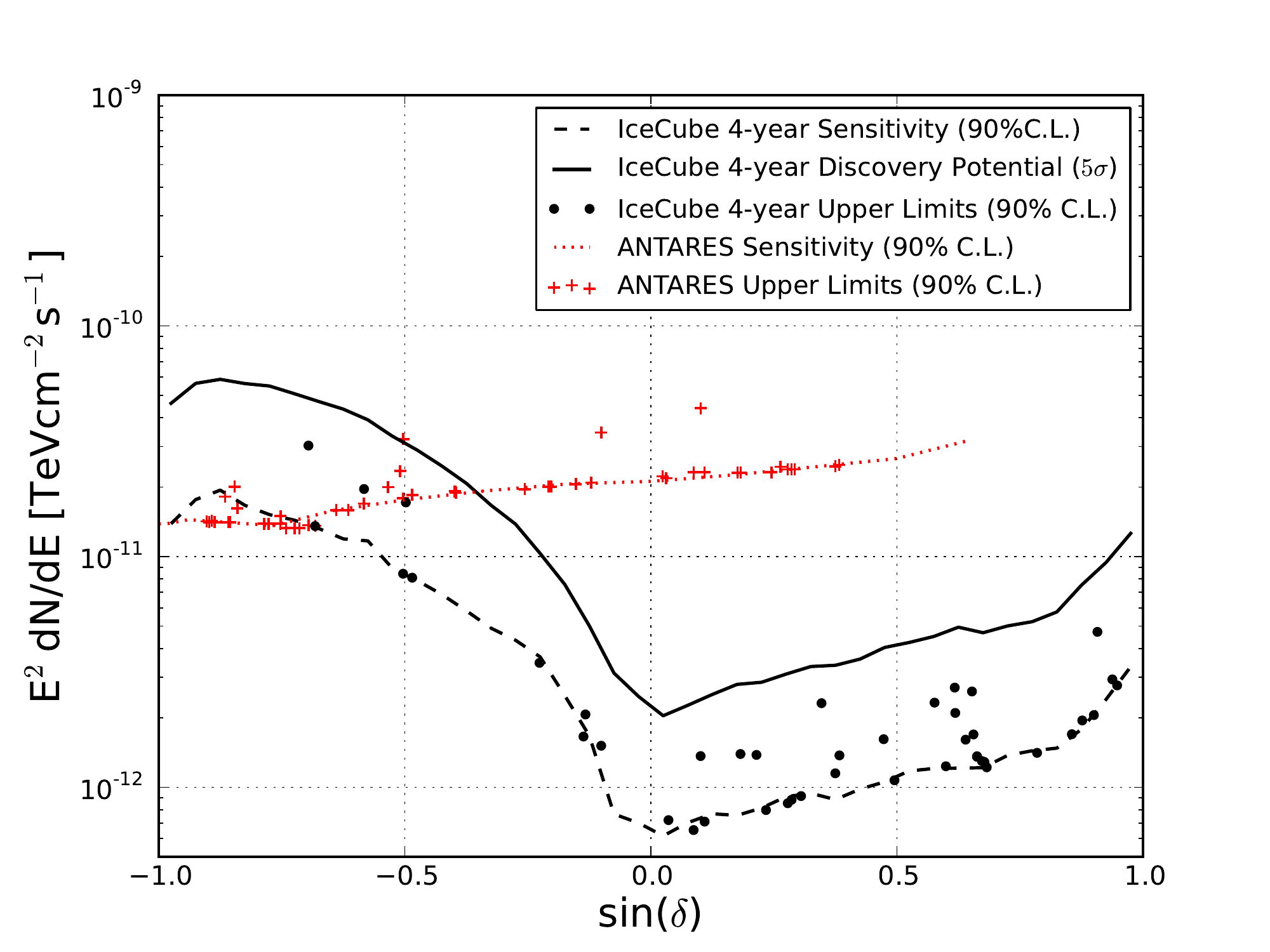}
\caption{Muon neutrino upper limits with 90\% C.L. evaluated for the 44 sources (dots), for the combined four years of data (40, 59, 79, and 86 string detector configurations). The solid black line is the flux required for $5\sigma$ discovery of a point source emitting an E$^{-2}$ flux at different declinations while the dashed line is the median upper limit or sensitivity also for a 90\% C.L.  The ANTARES sensitivities and upper limits are also shown~\citep{Antaref}.  For sources in the southern hemisphere, ANTARES constrains neutrino fluxes at lower energies than this work.}
  \label{fig:upperlimits}
\end{figure}

The search for neutrino emission from an \emph{a priori} list of 44 candidate sources produced the results shown in Tables \ref{tab:slGalactic} and \ref{tab:slExtraGalactic}.  In the northern sky, 1ES 0229+200 has the strongest upward fluctuation.  The pre-trial p-value of such a fluctuation is 0.053, but after considering the random chance of observing a fluctuation as strong or stronger than this in any of the sources, the post-trial p-value is 0.61.  In the southern sky, PKS 0537-441 has the strongest upward fluctuation, with a pre-trials p-value of 0.083 and a post-trials p-value of 0.33.  Upper limits on the E$^{-2}$ muon neutrino flux for 90\% confidence level (C.L.) from each source are listed in the table, and are shown along with the analysis sensitivity in Figure \ref{fig:upperlimits}.

\begin{figure}
\centering
\includegraphics[width=0.63\linewidth]{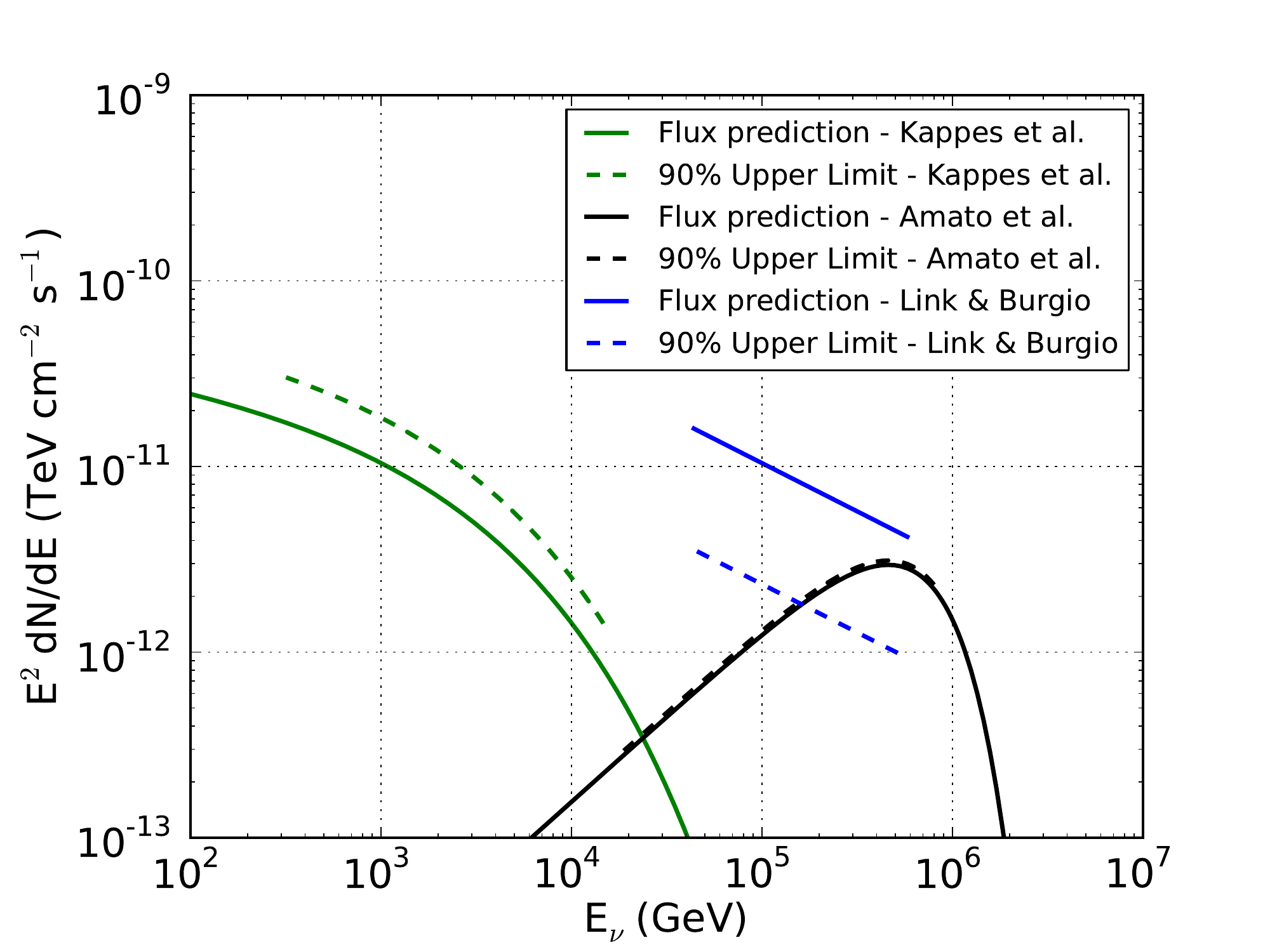}
\caption{Flux predictions (solid) for three models of neutrino emission from the Crab Nebula, with their associated 90\% C.L. upper limits (dashed) for an energy range containing 90\% of the signal.  Both the model from Amato {\it et al.}~\citep{Amato} and the most optimistic model from Link \& Burgio~\citep{LinkBurgio1, LinkBurgio2} are now excluded at 90\% confidence level.  For the gamma-ray based model from Kappes {\it et al.}~\citep{Kappes}, the upper limit is still a factor of 1.75 above the prediction.}
\label{fig:CrabUL}
\end{figure}

\begin{figure}
\centering
\includegraphics[width=0.63\linewidth]{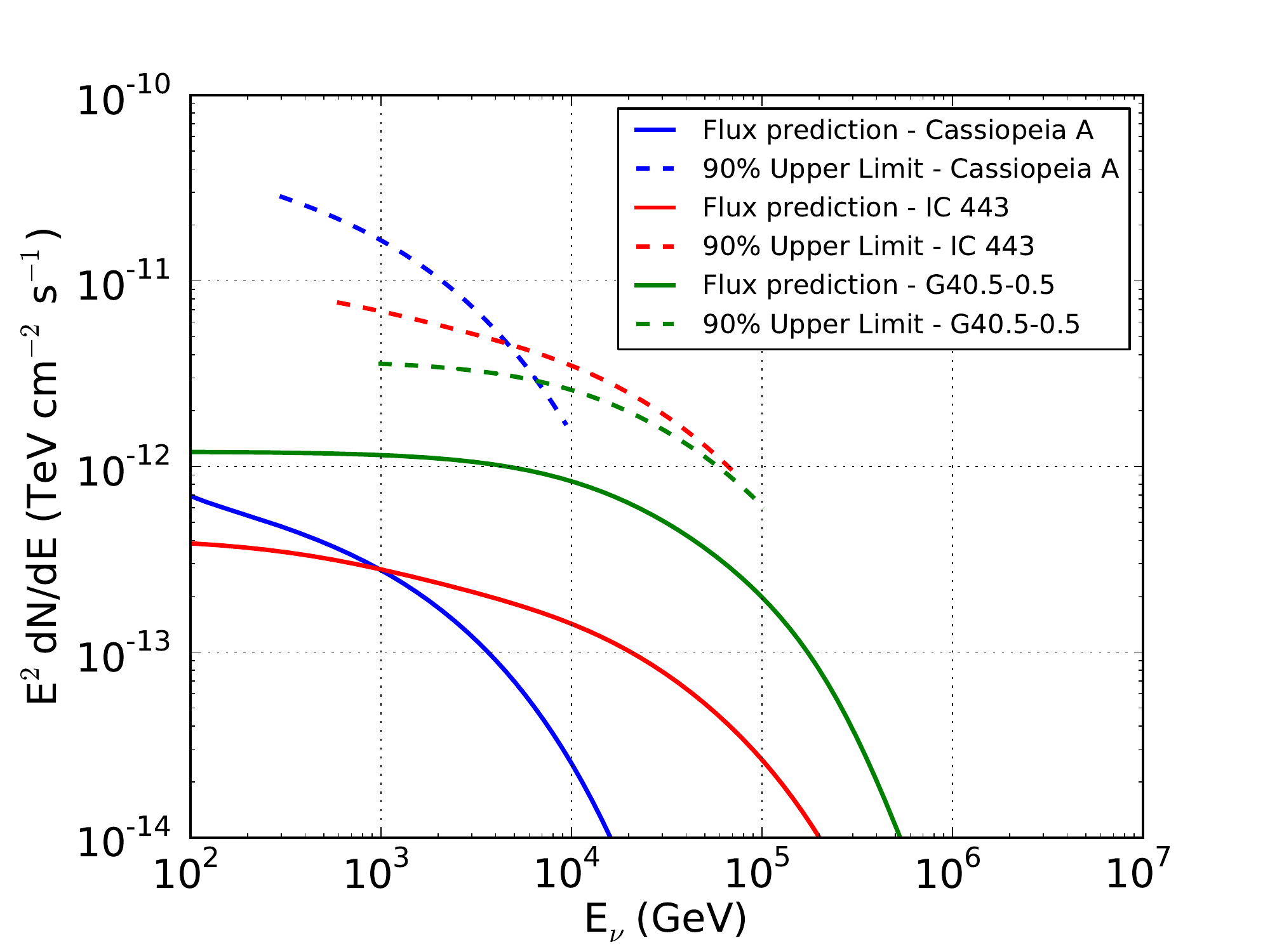}
\caption{Flux predictions (solid) and upper limits (dashed) for three Galactic supernova remnants.  The neutrino models, based of fitted gamma-ray observations, are from \citep{MCNew}.  For the source with the highest predicted flux, G40.5-0.5, the upper limit is a factor of three above the model.}
\label{fig:SNRUL}
\end{figure}

While many baseline models for CR acceleration and high-energy neutrino production predict E$^{-2}$ neutrino spectra, individual sources with unique conditions can produce significantly different spectra.  Models for any source in the sky can be tested with the analysis method used in this work, and a number of individual sources were previously considered in Ref.~\citep{IC79Paper}.  
Here, we update the 90\% C.L. upper limits on three models of neutrino emission from the Crab Nebula (Fig.~\ref{fig:CrabUL}) as well as three Galactic supernova remnants (Fig.~\ref{fig:SNRUL}).

\clearpage
 \begin{center}
 \vspace{0.5cm}
 \begin{longtable}[!t]{ p{1.7cm}  | l |r | r | c | c | c |c | c  }
 \caption{\label{tab:gul} Results for Galactic objects on the {\it a priori} search list. }\\
 \toprule
 Category & Source & r.a. [$^{\circ}$] & dec. [$^{\circ}$] & p-value & $\hat{n}_{S}$ & $\hat{\gamma}$  & B$_{1^{\circ}}$ & $\Phi_{\nu_{\mu} + \bar{\nu}_{\mu}}^{90\%}$ \\
 \midrule[\heavyrulewidth]
 \endfirsthead
 \multicolumn{9}{c}
 {\tablename\ \thetable\ -- \textit{Continued from previous page}} \\[{0.5cm}]
 \toprule
 Category & Source & r.a. [$^{\circ}$] & dec. [$^{\circ}$] & p-value & $\hat{n}_{S}$ & $\hat{\gamma}$ & B$_{1^{\circ}}$ & $\Phi_{\nu_{\mu} + \bar{\nu}_{\mu}}^{90\%}$ \\
 \midrule[\heavyrulewidth]
 \endhead
 \bottomrule \multicolumn{9}{r}{\textit{Continued on next page}} \\[{0.5cm}]
 \endfoot
 \hline
 \multicolumn{9}{l}{}\\[{0.1cm}]
 \multicolumn{9}{l}{\parbox{\textwidth}{\textit{ Note. -- } Sources are grouped according to their classification as High-Mass X-ray binaries or micro-quasars (HMXB/mqso),  SNRs, Pulsar Wind Nebulas (PWNs), star formation regions and unidentified sources. The p-value is the pre-trial probability of compatibility with the background-only hypothesis. The $\hat{n}_{S}$ and $\hat{\gamma}$ columns give the best-fit number of signal events and spectral index of a power-law spectrum. When $\hat{n}_{S} = 0$, no p-value or $\hat{\gamma}$ are reported.
 The eighth column gives the number of background events in a circle of 1$^{\circ}$ around the search coordinates. The last column shows the upper limits based on the classical approach \citep{Neyman} for an E$^{-2}$ flux normalization of $\nu_\mu + \bar{\nu}_{\mu}$ flux in units of $10^{-12}$\,TeV$^{-1}$cm$^{-2}$s$^{-1}$.\\
     }}\\[{1cm}]
 
 \endlastfoot

    {SNR} &  TYCHO &   6.36 & 64.18  & -- & 0.0 & -- & 17.8 & 2.06\\
       &         Cas A & 350.85 & 58.81 & -- & 0.0 & -- & 17.8 & 1.70 \\
       &         IC443 &  94.18 & 22.53  & 0.35 & 4.6 & 3.9 & 27.8 & 1.38\\
            \midrule         
   {HMXB} 
 &  LSI +63 303 &  40.13 & 61.23  & -- & 0.0 & -- & 17.8 & 1.95\\
 /mqso &     Cyg X-3 & 308.11 & 40.96 & 0.42 & 3.7 & 3.9 & 21.5 & 1.70 \\
    &          Cyg X-1 & 299.59 & 35.20  & 0.18 & 8.9 & 3.9 & 23.4 & 2.33\\
   & HESS J0632+057 &  98.25 &  5.80 & 0.14 & 13.4 & 3.4 & 37.0 & 1.37 \\
  &              SS433 & 287.96 &  4.98 & -- & 0.0 & -- & 37.6 & 0.65 \\
       \midrule
      {Star Formation Region} 
      &   Cyg OB2 & 308.08 & 41.51  & -- & 0.0 & -- & 21.0 & 1.36\\
 \midrule         
   {pulsar/} 
   &  MGRO J2019+37 & 305.22 & 36.83  & -- & 0.0 & -- & 23.1 & 1.23\\
 {PWN} & Crab Nebula &  83.63 & 22.01  & 0.44 & 4.4 & 3.9 & 27.8 & 1.15\\
 &  Geminga &  98.48 & 17.77  & -- & 0.0 & -- & 30.7 & 0.92\\
  \midrule
     {Galactic Center}
        &        Sgr A* & 266.42 & -29.01 & -- & 0.0 & -- & 36.6 & 8.11 \\
   %\midrule
   \newpage
    {Not identified} & MGRO J1908+06 & 286.98 & 6.27 &  -- & 0.0 & -- & 36.4 & 0.71 \\
    
 \label{tab:slGalactic}
 \end{longtable}
 \end{center}

 \begin{center}
 \vspace{-1cm}
 \begin{longtable}[th!]{ p{1.7cm}  | l |r | r | c | c | c |c | c  }
 \caption{\label{tab:eul} Results for extragalactic objects on the {\it a priori} search list. }\\
 \toprule
 Category & Source & r.a. [$^{\circ}$] & dec. [$^{\circ}$] & p-value & $\hat{n}_{S}$ & $\hat{\gamma}$  & B$_{1^{\circ}}$ & $\Phi_{\nu_{\mu} + \bar{\nu}_{\mu}}^{90\%}$ \\
 \midrule[\heavyrulewidth]
 \endfirsthead
 \multicolumn{9}{c}
 {\tablename\ \thetable\ -- \textit{Continued from previous page}} \\[{0.5cm}]
 \toprule
 Category & Source & r.a. [$^{\circ}$] & dec. [$^{\circ}$] & p-value & $\hat{n}_{S}$ & $\hat{\gamma}$ & B$_{1^{\circ}}$ & $\Phi_{\nu_{\mu} + \bar{\nu}_{\mu}}^{90\%}$ \\
 \midrule[\heavyrulewidth]
 \endhead
 \bottomrule \multicolumn{9}{r}{\textit{Continued on next page}} \\[{0.5cm}]
 \endfoot
 \hline
 \multicolumn{9}{l}{}\\[{0.1cm}]
 \multicolumn{9}{l}{\parbox{\textwidth}{\textit{ Note. -- }  Sources are grouped according to their classification as BL Lac objects, Radio Galaxies, Flat-Spectrum Radio Quasars (FSRQ) and Starburst galaxies. The p-value is the pre-trial probability of compatibility with the background-only hypothesis. The $\hat{n}_{S}$ and $\hat{\gamma}$ columns give the best-fit number of signal events and spectral index of a power-law spectrum. When $\hat{n}_{S} = 0$, no p-value or $\hat{\gamma}$ are reported.
 The eighth column gives the number of background events in a circle of 1$^{\circ}$ around the search coordinates. The last column shows the upper limits based on the classical approach \citep{Neyman} for an E$^{-2}$ flux normalization of $\nu_\mu + \bar{\nu}_{\mu}$ flux in units of $10^{-12}$\,TeV$^{-1}$ cm$^{-2}$s$^{-1}$.\\
     }}\\[{1cm}]
 \multicolumn{9}{l}{\parbox{\textwidth}{$^{a,b}$Most significant p-value in the northern and southern skies, respectively, among all Galactic and extragalactic objects on the {\it a priori} search list.}}\\
 
 \endlastfoot
 {BL Lac} & S5 0716+71 & 110.47 & 71.34 & -- & 0.0 & -- & 16.5 & 2.77 \\
 	      & 1ES 1959+650 & 300.00 & 65.15 & 0.083 & 9.8 & 3.2 & 17.7 & 4.72 \\
                &  1ES 2344+514 & 356.77 & 51.70 & -- & 0.0 & -- & 19.1 & 1.41 \\
 &  3C66A &  35.67 & 43.04  & -- & 0.0 & -- & 20.5 & 1.220\\
       &  H 1426+428 & 217.14 & 42.67  & -- & 0.0 & -- & 20.8 & 1.29\\
       &   BL Lac & 330.68 & 42.28 & -- & 0.0 & -- & 20.8 & 1.30 \\
        &  Mrk 501 & 253.47 & 39.76  & 0.45 & 3.2 & 3.7 & 22.1 & 1.61\\
        &  Mrk 421 & 166.11 & 38.21  & 0.26 & 3.8 & 1.9 & 22.4 & 2.10\\
        & W Comae & 185.38 & 28.23  & 0.34 & 1.4 & 1.6 & 25.9 & 1.62\\
        & 1ES 0229+200 &  38.20 & 20.29 & 0.053$^{a}$ & 16.0 & 3.7 & 28.6 & 2.32 \\
        & PKS 0235+164 &  39.66 & 16.62  & -- & 0.0 & -- & 31.4 & 0.88\\
        &  PKS 2155-304 & 329.72 & -30.23  & -- & 0.0 & -- & 37.0 & 8.43\\
        &   PKS 0537-441 &  84.71 & -44.09 & 0.083$^{b}$ & 6.3 & 3.9 & 35.2 & 30.03\\
 \midrule
 {FSRQ} 
 &             4C 38.41 & 248.81 & 38.13  & 0.12 & 10.6 & 2.8 & 22.4 & 2.71\\
 &         3C 454.3 & 343.49 & 16.15 & -- & 0.0 & -- & 31.4 & 0.85 \\
    &     PKS 0528+134 &  82.73 & 13.53 & -- & 0.0 & -- & 32.3 & 0.80 \\
  &         PKS 1502+106 & 226.10 & 10.49  & 0.21 & 6.1 & 2.3 & 33.2 & 1.39\\
  &             3C 273 & 187.28 &  2.05  & 0.45 & 3.2 & 2.6 & 38.9 & 0.72\\
   &             3C279 & 194.05 & -5.79 & -- & 0.0 & -- & 33.5 & 1.51 \\
    &     QSO 2022-077 & 306.42 & -7.64 & 0.45 & 1.3 & 2.0 & 34.1 & 2.07 \\
      &   PKS 1406-076 & 212.24 & -7.87  & -- & 0.0 & -- & 34.1 & 1.66\\
     &    QSO 1730-130 & 263.26 & -13.08& -- & 0.0 & -- & 37.1 & 3.46 \\
      & PKS 1622-297 & 246.53 & -29.86  & 0.13 & 6.2 & 2.7 & 36.6 & 17.17\\
      &    PKS 1454-354 & 224.36 & -35.65& 0.2 & 5.4 & 3.9 & 35.6 & 19.64 \\
  \midrule
     {Starburst}   &    M82 & 148.97 & 69.68 & -- & 0.0 & -- & 16.3 & 2.94 \\ 
 
 %\midrule
      {Radio}
   & NGC 1275 &  49.95 & 41.51 & -- & 0.0 & -- & 21.0 & 1.36 \\
 Galaxies &     Cyg A & 299.87 & 40.73 & 0.18 & 1.8 & 1.5 & 21.5 & 2.60 \\
        &    3C 123.0 &  69.27 & 29.67  & -- & 0.0 & -- & 25.7 & 1.07\\
        &  M87 & 187.71 & 12.39 & 0.26 & 8.8 & 3.9 & 32.4 & 1.38 \\
           &             Cen A & 201.37 & -43.02 & -- & 0.0 & -- & 35.5 & 13.57\\

 \label{tab:slExtraGalactic}
 \end{longtable}
\end{center}

\subsection{Stacking Searches}

\begin{figure}[!th]
  \vspace{5mm}
  \centering
  \includegraphics[width=.80\textwidth]{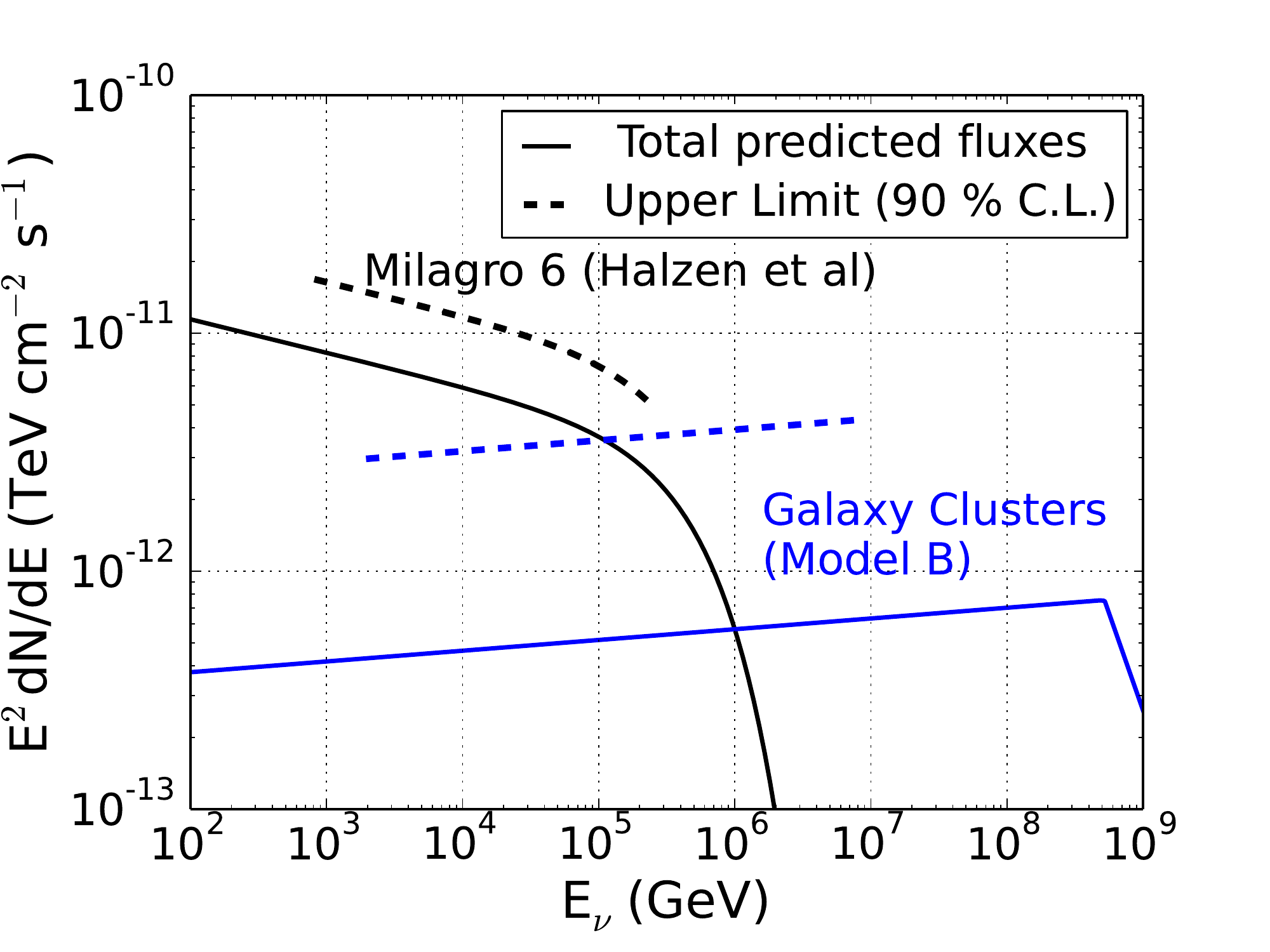}
  \caption{IceCube 90\% C.L. upper limits to the models of ~\citep{halzen2008} and ~\citep{galaxycluster}. }
  \label{fig:MilagroUpperLimits}
 \end{figure}

The results of all stacking searches are compatible with the background only hypothesis and are summarized in Table ~\ref{tab:Stacktable}. The most significant deviation from the background only hypothesis was observed in the stacked search for neutrino emission from the six Milagro TeV Gamma ray sources, with a p-value of 0.02. The fitted spectral index of 3.95 however suggests that only low energy events contribute towards the observation and the observed significance is from spatial clustering only. While Ref.~\citep{halzen2008} predicts a flux of much higher energy neutrinos from these sources, the assumptions made about the gamma ray spectra of the sources in Ref.~\citep{halzen2008} have later proved to be too optimistic~\citep{milagro2012}. Subsequently, the authors have updated the models~\citep{MilagroNewHalzenPaper}. Fig \ref{fig:MilagroUpperLimits} shows the IceCube upper limits to the model of Ref.~\citep{halzen2008}. In Fig \ref{fig:MilagroUpperLimits}, we also compare limits on neutrino fluxes from galaxy clusters to the model from Ref.~\citep{galaxycluster}.

\begin{table}[htbp]
 \centering
 \begin{tabular}[b]{ l  l |  c | c | c | c}
  \toprule
   \multicolumn{2}{l|}{Catalog}  &  $\hat{n}_{S}$ & $\hat{\gamma}$ & p-value & $\Phi_{\nu_{\mu} + \bar{\nu}_{\mu}}^{90\%}$\\
  \midrule[\heavyrulewidth]
    \multicolumn{2}{l|}{Milagro 6} & 51.4 & 3.95 & 0.02 & 1.98$\times$M.F. ~\citep{halzen2008}\\
   Galaxy Clusters & {\it Model A} & 1.4 & 3.95 & 0.50 & 3.89$\times$M.F. ~\citep{galaxycluster}\\
  & {\it Model B} & 12.6 & 3.95 & 0.48 & 6.17$\times$M.F.~\citep{galaxycluster}\\
  & {\it Central AGN} & 0.0  & --  & -- & 1.54$\times$M.F.~\citep{galaxycluster}\\
  & {\it Isobaric} &  0.0 & -- & -- & 4.65$\times$M.F.~\citep{galaxycluster}\\
  \multicolumn{2}{l|}{Starburst Galaxies} &  0.0 &--  & -- & $7.93\times10^{-12}$ $\times$E$^{2.0}$ \\
  \multicolumn{2}{l|}{MC Associated SNRs} &  0.0 & -- & -- & $1.60\times10^{-9}$ $\times$E$^{2.7}$ \\
  \multicolumn{2}{l|}{Supermassive Black Holes} & 17.1 & 3.95 & $0.43$ & $6.88\times10^{-12}$ $\times$E$^{2.0}$ \\
  \multicolumn{2}{l|}{Young SNRs} &  0.0 & -- & -- & $4.83\times10^{-12}$ $\times$E$^{2.0}$ \\
  \multicolumn{2}{l|}{Young PWNs} &  0.0 & --  & -- &  $3.12\times10^{-12}$ $\times$E$^{2.0}$\\
	
  FSRQs & W1 & 9.8 & 2.45 & 0.31 & 3.46 $\times10^{-12}\times$ E$^{2.0}$ \\
  & W2 & 15.4 & 2.75 & 0.19 & 34.3 $\times$ M.F. \\
 LSP BL Lacs & W1 & 11.9 & 3.25 & 0.38 & 5.24 $\times10^{-12}\times$ E$^{2.0}$\\
  & W2& 21.8 & 3.59 & 0.10 & 13.5 $\times$ M.F.\\
 Hard BL Lacs & W1 & 0     & --   & --    & 3.73 $\times10^{-12}\times$ E$^{2.0}$ \\
  & W2& 17.5 & 3.95 & 0.29 & 0.284 $\times$ M.F. \\
	
  \bottomrule
 \end{tabular}
\caption[Results of the stacked searches for emission from source catalogs]{Results of the stacked searches for emission from source catalogs. M.F. stands for the model flux as described in the references motivating the analyses. $\Phi_{\nu_{\mu} + \bar{\nu}_{\mu}}^{90\%}$ is the 90\% Confidence Level upper limit on the combined flux of $\nu_{\mu}$ and $\bar{\nu}_{\mu}$ from the catalogs. The E$^{2.0}$ limits are in units of ${\rm\,TeV}^{1}  {\rm cm}^{-2}  {\rm s}^{-1}$.}
\label{tab:Stacktable}
\end{table}

\subsection{Systematic Uncertainties}

In all analyses described here the background is estimated by scrambling the detector data in right ascension and is independent of theoretical uncertainties on fluxes of atmospheric neutrino and muons as well as uncertainties in the simulation of the detector.  The p-values are therefore robust against most sources of systematic error.  Upper limits and analysis sensitivities however are calculated by simulating the detector response to neutrinos.  Detector uncertainties including the optical properties of the ice and the absolute efficiency of the optical modules can affect the reported sensitivities and upper limits.

After a detailed discussion of all relevant systematic uncertainties, Ref.~\citep{IC79Paper} concludes that the level of uncertainty in the analysis using three years of data is about 18\%.  Since 65\% of the data used here is the same as in Ref.~\citep{IC79Paper} and the techniques for the new event selection and analyses are similar, the systematic uncertainty on the 4 year sample is about the same.  However, the added year of data utilizes a new muon track reconstruction, which is more sensitive to uncertainties in the optical properties of the ice.  We re-evaluate the effect of the ice properties on the analysis for the 2011-2012 data, finding a corresponding systematic uncertainty of +16\%/-8\%.  This is incorporated into the overall systematic uncertainty by averaging it with the ice model effect from the previous years.  The resulting overall systematic uncertainty on the quoted sensitivities and upper limits is 21\%.

\section{Conclusions}
\label{sec7}

\begin{figure}
\centering
\includegraphics[width=0.75\textwidth]{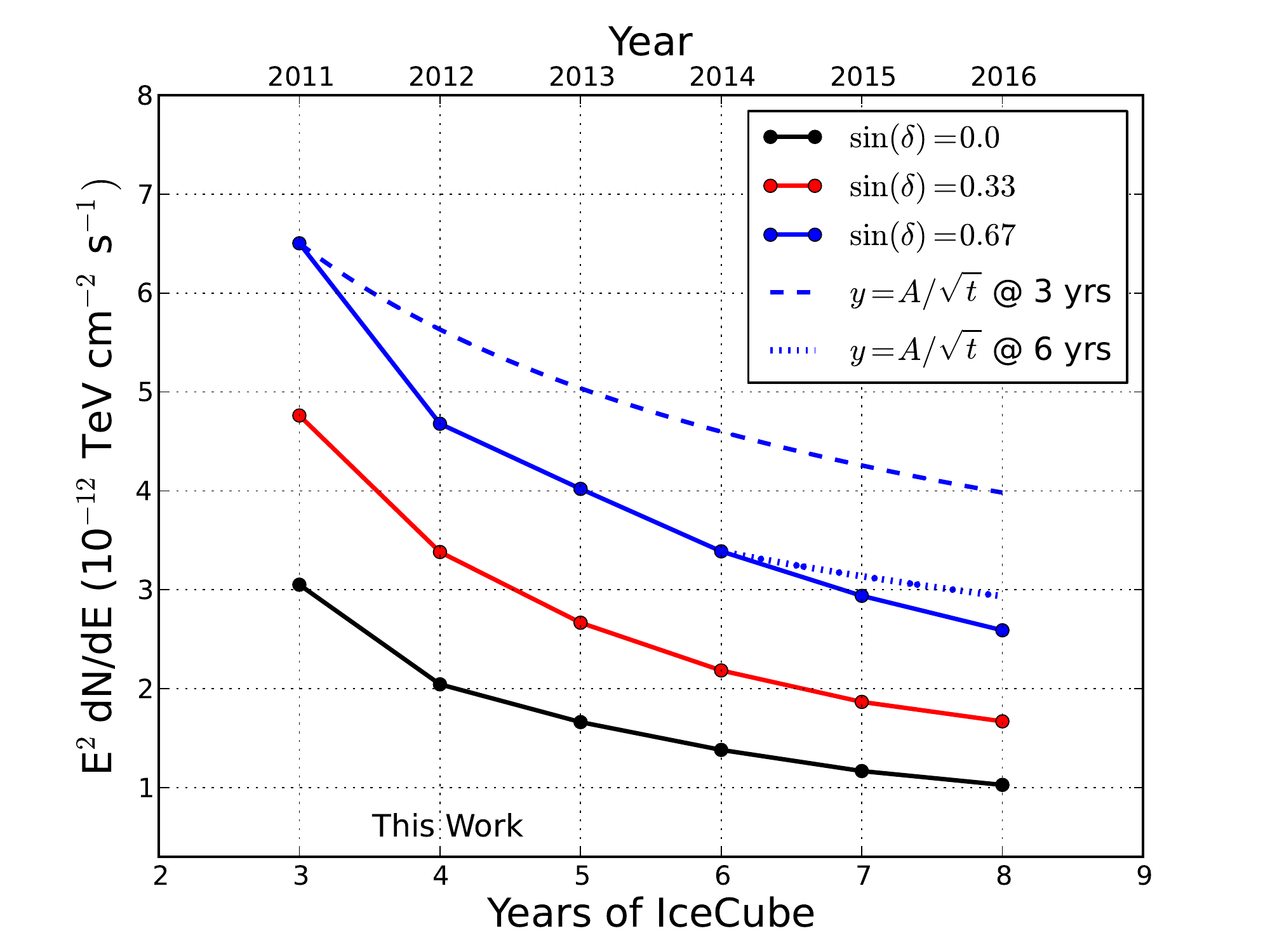}
\caption{Predicted $E^{-2}$ discovery potential as a function of years of running time of the IceCube Observatory for three different declinations (solid lines).  Due to the relatively low background rate in this analysis, the discovery potential will continue to improve faster than the square-root of time limit (dashed, dotted lines).}
\label{fig:DiscoVsTime}
\end{figure}

No evidence of neutrino emission from point-like or extended sources was found in four years of IceCube data.  Searches for emissions from point-like and extended sources anywhere in the sky, from a pre-defined candidate source list and from stacked source catalogs all returned results consistent with the background-only hypothesis.  90\% C.L. upper limits on the muon neutrino fluxes for models from a variety of sources were calculated and compared to predictions. The most optimistic models considered here can be excluded at 90\% C.L. and in other cases limits are a factor of two to four above the predictions.
This analysis includes data from the completed IceCube array, taken between May 2011 and May 2012.  IceCube will continue to run in this configuration for the foreseeable future.  Future analyses will benefit from this improved integration time and the evolution of the analysis sensitivity as a function of years of data-taking is shown in Figure \ref{fig:DiscoVsTime}.  Within a few years the analyses will surpass the sensitivity necessary to test a wider variety of neutrino point source models.  Future developments in background rejection techniques and reconstruction algorithms may lead to improvements faster than predicted in Figure \ref{fig:DiscoVsTime}.

\section*{Acknowledgements}

We acknowledge the support from the following agencies:
U.S. National Science Foundation-Office of Polar Programs,
U.S. National Science Foundation-Physics Division,
University of Wisconsin Alumni Research Foundation,
the Grid Laboratory Of Wisconsin (GLOW) grid infrastructure at the University of Wisconsin - Madison, the Open Science Grid (OSG) grid infrastructure;
U.S. Department of Energy, and National Energy Research Scientific Computing Center,
the Louisiana Optical Network Initiative (LONI) grid computing resources;
Natural Sciences and Engineering Research Council of Canada,
WestGrid and Compute/Calcul Canada;
Swedish Research Council,
Swedish Polar Research Secretariat,
Swedish National Infrastructure for Computing (SNIC),
and Knut and Alice Wallenberg Foundation, Sweden;
German Ministry for Education and Research (BMBF),
Deutsche Forschungsgemeinschaft (DFG),
Helmholtz Alliance for Astroparticle Physics (HAP),
Research Department of Plasmas with Complex Interactions (Bochum), Germany;
Fund for Scientific Research (FNRS-FWO),
FWO Odysseus programme,
Flanders Institute to encourage scientific and technological research in industry (IWT),
Belgian Federal Science Policy Office (Belspo);
University of Oxford, United Kingdom;
Marsden Fund, New Zealand;
Australian Research Council;
Japan Society for Promotion of Science (JSPS);
the Swiss National Science Foundation (SNSF), Switzerland;
National Research Foundation of Korea (NRF);
Danish National Research Foundation, Denmark (DNRF)

%Bibliography


\begin{thebibliography}{}
\bibitem[Aartsen {\it et al.}(2013a)]{SPICEPaper}
M.~G.~Aartsen {\it et al.} [IceCube Coll.]  Nucl. Instr. Meth. {\bf A711} (2013) 73.
\bibitem[Aartsen {\it et al.}(2013b)]{ImprovedLineFitPaper}
M.~G.~Aartsen {\it et al.} [IceCube Coll.]  Nucl. Instr. Meth. {\bf A736} (2013) 143.
\bibitem[Aartsen {\it et al.}(2013c)]{IC79Paper}
M.~G.~Aartsen {\it et al.} [IceCube Coll.]  ApJ {\bf779} (2013) 132.
\bibitem[Aartsen {\it et al.}(2013d)]{HESEPaper}
M.~G.~Aartsen {\it et al.} [IceCube Coll.] Science {\bf 342} (2013) 947.
\bibitem[Aartsen {\it et al.}(2013e)]{moonpaper}
M.~G.~Aartsen {\it et al.} [IceCube Coll.] (2013) arXiv:1305.6811.
\bibitem[Aartsen {\it et al.}(2014a)]{EnergyRecoPaper}
M.~G.~Aartsen {\it et al.} [IceCube Coll.] JINST {\bf 9} (2014) P03009.
\bibitem[Aartsen {\it et al.}(2014b)]{HESE3Year}
M.~G.~Aartsen {\it et al.} [IceCube Coll.] (2014) arXiv 1405.5303.
\bibitem[Abbasi {\it et al.}(2009)]{DOMMBPaper}
R.~Abbasi {\it et al.} [IceCube Coll.] Nucl. Instr. Meth. {\bf A601} (2009) 294.
\bibitem[Abbasi {\it et al.}(2010)]{PMTPaper}
R.~Abbasi {\it et al.} [IceCube Coll.] Nucl. Instr. Meth. {\bf A618} (2010) 139.
\bibitem[Abbasi {\it et al.}(2011)]{IC40Paper} 
R.~Abbasi {\it et al.} [IceCube Coll.], ApJ {\bf 732} (2011) 18.
\bibitem[Abdo {\it et al.}(2007)]{milagro2007}
A.~A.~Abdo {\it et al.}, ApJ {\bf 664} (2007) L91.
\bibitem[Abdo {\it et al.}(2009a)]{W51CFermi}
A.~A.~Abdo {\it et al.} [Fermi Coll.], ApJ {\bf 706} (2009) L1.
\bibitem[Abdo {\it et al.}(2009b)]{W51CMilagro}
A.~A.~Abdo {\it et al.} [Milagro Coll.], ApJ. {\bf 700} (2009) L127.
\bibitem[Abdo {\it et al.}(2010)]{W44Fermi}
A.~A.~Abdo {\it et al.} [Fermi Coll.], Science {\bf 327} (2010) 1103.
\bibitem[Abdo {\it et al.}(2012)]{milagro2012}
A.~A.~Abdo {\it et al.}, ApJ {\bf 753} (2012) 159. 
\bibitem[Achterberg {\it et al.}(2006)]{FirstYearPerformancePaper}
A.~Achterberg {\it et al.}, [IceCube Coll.] Astropart. Phys. {\bf 26} (2006) 155.
\bibitem[Ackermann {\it et al.}(2011)]{FermiAGNCat}
M.~Ackermann {\it et al.}, ApJ {\bf743} (2011) 171.
\bibitem[Ackermann {\it et al.}(2013)]{FermiSNR}
M.~Ackermann {\it et al.} [Fermi Coll.], Science {\bf 339} (2013) 807.
\bibitem[Adri\'{a}n-Mart\'{i}nez {\it et al.}(2014)]{Antaref}
S.~Adri\'{a}n-Mart\'{i}nez {\it et al.} [Antares Coll.] ApJ {\bf 786} (2014) L5.
\bibitem[Ahrens {\it et al.}(2004)]{AMANDAMuonRecoPaper}
J.~Ahrens {\it et al.} Nucl. Instrum. Meth. {\bf A524} (2004) 169.
\bibitem[Alvarez-Mu{\~n}iz \& Halzen(2002)]{2002ApJL}
J.~Alvarez-Mu{\~n}iz and F.~Halzen, ApJ {\bf 576} (2002)  L33.
\bibitem[Amato {\it et al.} (2003)]{Amato}
E.~Amato, D.~Guetta and P.~Blasi, Astron. Astrophys. {\bf 402} (2003) 827.
\bibitem[Anchordoqui \& Montaruli(2010)]{AnchorMonta}
L.~Anchordoqui and T.~Montaruli, Ann. Rev. Nucl. Part. Sci. {\bf 60} (2010) 129.
\bibitem[Anchordoqui {\it et al.}(2014)]{AnchorReview}
L.~Anchordoqui {\it et al.} JHEAp. {\bf 1} (2014) 1.
\bibitem[Atoyan \& Dermer(2001)]{DermerModelPaper}
A.~Atoyan \& C.D. Dermer, Phys. Rev. Lett. {\bf87} (2001) 221102.
\bibitem[Becker (2008)]{2008Becker}
J.~Becker, Phys. Rep. {\bf 458} (2008) 173.
\bibitem[Becker {\it et al.}(2009)]{Starburst}
J.~Becker {\it et al.}, (2009) arXiv:0901.1775.
\bibitem[Bednarek(2003)]{PWNPaper}
W.~Bednarek, Astron. Astrophys. {\bf 407 } (2003) 1.
\bibitem[Braun {\it et al.}(2010)]{method}
J.~Braun {\it et al.}, Astropart. Phys.  {\bf 33} (2010) 175.
\bibitem[Caramete \& Biermann(2010)]{bh}
L.~Caramete \& P.L. Biermann, Astron. Astrophys. {\bf 521} (2010) A55.
\bibitem[Carrigan {\it et al.}(2013)]{HESSsurvey}
S.~Carrigan {\it et al.} [HESS Coll.] (2013) arXiv:1307.4868v2.
\bibitem[Castro {\it et al.}(2011)]{Castro}
D.~Castro {\it et al.}, ApJ {\bf 734} (2011) 85.
\bibitem[Cavasinni {\it et al.}(2006)]{MC}
V.~Cavasinni {\it et al.}, Astropart. Phys. {\bf 26} (2006) 41.
\bibitem[De Marco {\it et al.}(2006)]{2006PRD}
D~.De Marco, P.~Blasi, P.~Hansen, and T.~Stanev. Phys. Rev. {\bf D73} (2006) 043004.
\bibitem[Essey {\it et al.}(2010)]{2010PRL}
W.~Essey, O.~E.~Kalashev, A.~Kusenko, and J.~F.~Beacom, Phys. Rev. Lett. {\bf 104} (2010) 141102.
\bibitem[Ferrand \& Safi-Harb(2012)]{SNRCat}
G.~Ferrand and S.~Safi-Harb, Advances in Space Research {\bf 49} (2012) 1313.
\bibitem[Fiasson {\it et al.}(2009)]{W51CHess}
A.~Fiasson {\it et al.} [HESS Coll.], in {\it Proc. of the } 31$^{st}$ Int. Cosmic Ray Conf., Lodz, Poland (2009).
\bibitem[Gonzalez-Garcia {\it et al.}(2014)]{MilagroNewHalzenPaper}
M.~C.~Gonzalez-Garcia {\it et al.} Astrop. Phys. {\bf 57} (2014) 39.
\bibitem[Greisen(1966)]{GZK}
K.~Greisen, Phys. Rev. Lett. {\bf 16} (1966) 748.
\bibitem[Guetta {\it et al.}(2004)]{Guetta}
D.~Guetta {\it et al.} Astrop. Phys. {\bf 20} (2004) 429.
\bibitem[Halzen \& Hooper(2002)]{2002HalzenHooper}
F.~Halzen and D.~Hooper, Rep. Progr. Phys. {\bf 65} (2002) 1025.
\bibitem[Halzen {\it et al.}(2008)]{halzen2008}
F.~Halzen, A.~Kappes and A.~O'Murchadha, Phys. Rev. {\bf D78} (2008) 063004.
\bibitem[Honda {\it et al.}(2007)]{Honda}
M.~Honda {\it et al.}, Phys.Rev. {\bf D75} (2007) 043006.
\bibitem[Kalashev {\it et al.}(2013)]{2013PRL}
O.~E.~Kalashev, A.~Kusenko, and W.~Essey, Phys. Rev. Lett. {\bf 111} (2013) 041103.
\bibitem[Kappes {\it et al.}(2007)]{Kappes}
A. Kappes {\it et al.}, ApJ {\bf 656} (2007) 870.
\bibitem[Kistler \& Beacom(2006)]{2006PRD_KB}
M.~D.~Kistler \& J.~F.~Beacom, Phys. Rev. {\bf D74} (2006) 0607082.
\bibitem[Lacki {\it et al.}(2011)]{2011ApJStarbursts}
B.~C.~Lacki {\it et al.}, ApJ {\bf 734} (2011) 107.
\bibitem[Learned \& Mannheim(2000)]{2000LearnedMannheim}
J.~G.~Learned and K.~Mannheim, Ann. Rev. Nucl. Part. Sci. {\bf 50} (2000) 679.
\bibitem[Link \& Burgio(2005)]{LinkBurgio1}
B. Link \& F. Burgio, Phys. Rev. Lett. {\bf 94} (2005) 181101.
\bibitem[Link \& Burgio(2006)]{LinkBurgio2}
B. Link \& F. Burgio, Mon. Not. Roy. Astron. Soc. {\bf 371} (2006) 375.
\bibitem[Loeb \& Waxman(2006)]{LoebWaxman}
A.~Loeb and E.~Waxman, JCAP {\bf 0605} (2006) 003.
\bibitem[Mandelartz \& Tjus(2013)]{MCNew}
M.~Mandelartz \& J.~Becker Tjus, (2013) arxiv:1301.2437.
\bibitem[M{\'e}sz{\'a}ros(2006)]{Meszaros}
P.~M{\'e}sz{\'a}ros, Rep. Prog. Phys. {\bf 69} (2006) 2259.
\bibitem[M\"ucke {\it et al.}(2003)]{MueckeModelPaper}
A.~M$\mathrm{\ddot{u}}$cke {\it et al.}, Astropart.Phys. {\bf18} (2003) 593.
\bibitem[Murase {\it et al.}(2008)]{galaxycluster}
K.~Murase {\it et al.}, ApJ {\bf 689} (2008) L105.
\bibitem[Murase \& Beacom(2012)]{MuraseClusters}
K.~Murase and J.~F.~Beacom, JCAP {\bf 02} (2012) 028.
\bibitem[Murase {\it et al.}(2013)]{MuraseStarbursts}
K.~Murase, M.~Ahlers, and B.~C.~Lacki, Phys. Rev. {\bf D88} (2013) 121301.
\bibitem[Murase {\it et al.}(2014)]{MuraseAGN}
K.~Murase, Y.~Inoue, and C.~D.~Dermer, (2014) arXiv:1403.4089.
\bibitem[Neronov \& Ribordy(2009)]{NeronovModelPaper}
A.~Neronov \& M. Ribordy, Phys.Rev. {\bf D80} (2009) 083008.
\bibitem[Neunh\"offer(2006)]{ParaboloidPaper}
T.~Neunh\"offer, Astropart. Phys. {\bf 26} (2006) 220.
\bibitem[Neyman(1937)]{Neyman}
J. Neyman. Phil. Trans. Royal Soc. London A {\bf 236} (1937) 333.
\bibitem[Romero \& Torres(2003)]{2003ApJStarbursts}
G.~E.~Romero and D.~F.~Torres, ApJ {\bf 586} (2003) L33.
\bibitem[Sironi \& Spitkovsky(2011)]{2011ApJ}
L.~Sironi and A.~Spitkovsky, ApJ {\bf 726} (2011) 75.
\bibitem[Sedov(1946)]{Sedov}
L.~I.~Sedov, Journal of Applied Mathematics and Mechanics, Vol. {\bf 10} (1946) 241.
\bibitem[Stecker {\it et al.}(1991)]{1991Stecker}
F.~W.~Stecker, C.~Done, M.~H.~Salamon, and P.~Sommers, Phys. Rev. Lett. {\bf 66} (1991) 2697.
\bibitem[Tchernin {\it et al.}(2013)]{NeronovTeresaExtendedPaper}
C.~Tchernin {\it et al.}, Astron. Astrophys. {\bf 560} (2013) A67.
\bibitem[Vissani {\it et al.}(2011)]{2011AstroPart}
F.~Vissani, F.~Aharonian, and N.~Sahakyan, Astrop. Phys. {\bf 34} (2011) 778.
\bibitem[Waxman \& Bahcall(1997)]{WaxmanBahcallGRB}
E.~Waxman and J.~Bahcall, Phys. Rev. Lett. {\bf 78} (1997) 2292.
\bibitem[Waxman \& Bahcall(1999)]{WaxmanBahcall}
E.~Waxman and J.~Bahcall, Phys. Rev. {\bf D59} (1999) 023002.
\bibitem[Whitehorn {\it et al.}(2013)]{PhotosplinePaper}
N.~Whitehorn, J.~van Santen and S.~Lafebre, CPC {\bf 184} (2013), 2214-2220.
\bibitem[Wolfe {\it et al.}(2008)]{2008ApJClusters}
B.~Wolfe, F.~Melia, R.~M.~Crocker, and R.~R.~Volkas, ApJ {\bf 687} (2008) 193.

\end{thebibliography}
\end{document}